\documentclass[fleqn,usenatbib]{mnras}

\usepackage{newtxtext,newtxmath}

\usepackage[T1]{fontenc}

\DeclareRobustCommand{\VAN}[3]{#2}
\let\VANthebibliography\thebibliography
\def\thebibliography{\DeclareRobustCommand{\VAN}[3]{##3}\VANthebibliography}

\usepackage{graphicx}	
\usepackage{amsmath}	
\usepackage{float}
\usepackage{fontawesome5}
\usepackage{hyperref}
\usepackage[all]{hypcap}	

\defcitealias{Sola2025}{S25}
\defcitealias{Chemaly2026}{C26}

\title[Constraining Dark Matter halo population]{Constraints on the population level distribution of nearby Dark Matter halo shapes with extragalactic streams}

\author[D. Chemaly et al.]{
David Chemaly$^{1}$\thanks{E-mail: dc824@cam.ac.uk},
Elisabeth Sola$^{1}$,
Sergey E. Koposov$^{2, 1}$,
HanYuan Zhang$^{1}$,
Vasily Belokurov$^{1}$,
\newauthor Denis Erkal$^{3}$
\\
$^{1}$ Institute of Astronomy, Madingley Rd, Cambridge CB3 0HA, UK\\
$^{2}$ Institute for Astronomy, University of Edinburgh, Royal
Observatory, Blackford Hill, Edinburgh EH9 3HJ, UK\\
$^{3}$Department of Physics, University of Surrey, Guildford, GU2 7XH, Surrey, UK\\
}

\date{Accepted XXX. Received YYY; in original form ZZZ}

\pubyear{\the\year{}}

\begin{document}
\label{firstpage}
\pagerange{\pageref{firstpage}--\pageref{lastpage}}
\maketitle

\begin{abstract}
Stellar streams trace the gravitational potential of their host galaxies and provide a sensitive probe of dark matter halo structure. Previously, we developed, and tested on simulated data, a hierarchical Bayesian framework to infer the population level distribution of dark matter halo shapes from ensembles of extragalactic stellar streams with images only. In this work, we apply this pipeline to 32 stellar streams from the \textsc{STRRINGS} catalogue, a curated sample of dynamically cold minor-merger streams detected in deep imaging. Each stream is forward-modelled assuming an axisymmetric halo and fitted using only the projected stream track, yielding posterior constraints on the halo flattening parameter $q$. To account for model mismatch and track systematics, we introduce an additional variance term that inflates the uncertainty on the projected stream track and use it to identify a high quality (\textit{gold}) subsample of 17 streams whose tracks retain significant constraining power. We then combine the individual posteriors through importance sampling to infer the underlying population distribution of halo flattening. For the \textit{gold} subsample, we infer an oblate population with mean $\mu_q \approx 0.72$ and intrinsic scatter $\sigma_q \approx 0.34$. Streams dominated by additional model variance yield a nearly spherical population inference. The inferred oblate population for the \textit{gold} sample is broadly consistent with expectations from cosmological hydrodynamical simulations. This work provides constraints on dark matter halo flattening from stellar streams beyond the Local Group and establishes a scalable framework for forthcoming large samples from \textit{Euclid} and Rubin/LSST.
\end{abstract}




\section{Introduction}

Galaxies assemble hierarchically through the accretion and merger of smaller systems \citep{White1978,Cole2000,Springel2006}. A natural outcome of this process is the formation of tidal debris: low surface brightness (LSB) structures produced as stars and gas are stripped from their galaxy during gravitational interactions. The morphology of this debris depends on the interaction channel. Minor mergers produce stellar streams: faint, elongated streams of stars created as dwarf galaxies or globular clusters are progressively stripped by the tidal field of their host \citep{Johnston1995,Helmi1999}. Because streams approximately trace the orbit of their progenitor, their geometry encodes information about the host potential and, by extension, the distribution of dark matter (DM) \citep{Johnston1998,Ibata2021,Johnston2005,Law2010,Koposov2010,Erkal2016,Bovy2016,Bonaca2018}. Streams simultaneously provide a record of late time accretion and a dynamical probe of halo structure linking galaxy assembly to the physics of DM.

In the $\Lambda$CDM paradigm, DM haloes are intrinsically non spherical. Collisionless $N$-body (i.e. DM only) simulations robustly find triaxial halo shapes whose axis ratios depend on halo mass, radius, redshift, and assembly history \citep{Frenk1988,Dubinski1991,Jing2002,Allgood2006,Bett2007,Despali2014,Bonamigo2015}. Haloes tend to be more prolate at early times and become increasingly spherical towards lower redshift due to ongoing accretion, violent relaxation, and phase mixing \citep{Allgood2006,Zemp2012,VegaFerrero2017}. The shape is also radially dependent, with inner regions generally rounder and outer regions more elongated \citep{Kazantzidis2004,Hayashi2007,VeraCiro2011}. Even at $z \sim 0$, DM haloes in collisionless simulations retain a mild prolate bias in the absence of baryonic effects \citep{Allgood2006,Despali2014,Bonamigo2015}.

Cosmological hydrodynamical simulations demonstrate that baryonic physics can substantially reshape DM haloes relative to collisionless predictions. Gas cooling deepens the central potential, while disk formation and time-dependent baryonic structures drive the DM distribution towards rounder and often oblate configurations aligned with the stellar disk \citep{Kazantzidis2004,Debattista2008,Abadi2010,Tissera2010}. This isotropisation is most pronounced in the inner halo, typically within the radii probed by stellar streams ($\leq100~\text{kpc}$), while the outer halo generally retains a triaxial or mildly prolate shape \citep{VeraCiro2011,Zemp2012}. This qualitative behaviour is consistently observed across modern cosmological hydrodynamical simulations, including \textit{Illustris} and \textit{IllustrisTNG} \citep{Chua2019}, \textit{EAGLE} \citep{Schaller2015,Velliscig2015}, \textit{FIRE} \citep{Hopkins2018,GarrisonKimmel2017}, and \textit{Auriga} \citep{Grand2017}. However, these simulations do not yield a single universal prediction for halo flattening at $z \sim 0$. Using 30 halos in \textit{Auriga}, \cite{Prada2019} obtained a minor to major axis ratio of roughly $0.72\pm0.9$ for a radial range $\leq 100 \text{kpc}$. The degree of halo rounding depends sensitively on galaxy mass, star formation efficiency, feedback implementation, and radial scale, leading to substantial scatter in the predicted distribution of axis ratios \citep{Chua2019,Prada2019}. Consequently, while baryons generically reduce triaxiality, the population level distribution of halo shapes at low redshift remains an open, empirically testable outcome of galaxy formation models.

Observationally, the Milky Way (MW) provides the most detailed and stringent constraints on DM halo shape, owing to the availability of six dimensional phase space information for multiple stellar streams. Analyses combining photometry, distances, proper motions, and radial velocities for streams such as Sagittarius, GD-1, Palomar~5, and Orphan–Chenab have enabled precise reconstructions of the Galactic potential and its substructure \citep{Law2010,Koposov2010,Bovy2016,Malhan2019,Erkal2019,Vasiliev2021,Koposov2023}. Collectively, these studies generally favour a halo that is close to spherical or mildly oblate in the inner regions, with increasing flattening or triaxiality at larger radii \citep{Law2010,VeraCiro2013,Bonaca2014,Bovy2016,Palau2023}. 

Despite this progress, MW halo shape measurements remain model dependent and exhibit significant scatter across tracers and methodologies, reflecting sensitivities to stream selection, assumed potential parametrisations, and the treatment of time-dependent perturbers \citep{VeraCiro2013,Erkal2019,Malhan2019,Vasiliev2021}. In particular, the influence of the Large Magellanic Cloud (LMC) has been shown to induce systematic biases if neglected, further complicating attempts to infer a single global halo morphology from individual streams \citep{Law2010,Erkal2019,GaravitoCamargo2021,Lilleengen2023}. These limitations underscore that, while the MW provides valuable constraints, it represents only one realisation of galaxy formation.

Extending stream-based halo shape measurements beyond the Local Group is therefore a crucial next step. Measuring halo morphologies across a diverse population of external galaxies would enable population level tests of structure formation, baryonic effects, and alternative DM scenarios that cannot be robustly addressed using a single system \citep{Jing2002,Allgood2006,Despali2014,Bonamigo2015,Chua2019,Prada2019}. In practice, however, streams around external galaxies are typically detected only as diffuse LSB features, without resolved stars or kinematic information. This loss of phase space dimensionality leads to strong degeneracies when modelling individual systems \citep{Johnston2001,Amorisco2015,Pearson2022}. However, photometry alone still holds constraining power: purely geometric observables encoded in the projected stream, namely curvature, orbital precession, and their radial dependence, have been shown to retain sensitivity to halo shape and to the radial structure of the potential \citep{Nibauer2023,Walder2024,Nibauer2025b}. These geometric methods have recently been applied to real extragalactic systems: \citet{Wu2026} inferred halo shapes from stellar-stream tracks in the local Universe, and \citet{Starkman2026} reconstructed the projected host potentials of 13 streams in the \textit{Euclid} Quick Data Release. The accelerating growth of deep, wide-field and large-scale imaging and the advent of new facilities, such as Euclid \citep{Euclid2022} and Rubin/LSST \citep{LSST2009}, will substantially increase the number of detectable LSB streams \citep{Laureijs2011,Ivezic2019,Martin2022}, further strengthening the case for statistical approaches that extract robust constraints from photometry-only data by coherently combining information from ensembles of streams.

A major practical limitation has been the scarcity of large, homogeneous samples of stellar streams that are both reliably detectable and well suited to dynamical modelling. Stream detectability depends strongly on imaging depth, background subtraction, and processing choices but also intrinsic parameters such as  orientation, time since merger, distance from host and the redshift. Therefore, the identification and classification of tidal features is not uniform across studies \citep{Atkinson2013,Hood2018}. As a result, many reported candidates are difficult to model robustly, for example because they are too short, morphologically ambiguous or embedded in systems with complex recent interactions. To help overcome this bottleneck, \citet{Sola2025} (hereafter \citetalias{Sola2025}) presented \textsc{STRRINGS} (\textit{STReams in Residual Images of Nearby GalaxieS}), a curated catalogue of long, narrow streams around nearby galaxies with a dynamically simple minor merger selected for forward modelling. \textsc{STRRINGS} was constructed by visually inspecting residual images from the DESI Legacy Imaging Surveys \citep{Dey2019}, where subtracting modelled light profiles enhances faint asymmetries and improves stream detectability \citep[e.g.][]{Bell2006,McIntosh2008,Tal2009}. From a parent sample of $19{,}387$ galaxies at $z \le 0.02$, \textsc{STRRINGS} provides a set of 35 streams together with segmentation products and recovered properties designed to interface with forward models, enabling consistent likelihood evaluation across the full observational sample.

In a companion paper, \citet{Chemaly2026} (hereafter \citetalias{Chemaly2026})
 developed a new hierarchical Bayesian framework designed to infer the population distribution of DM halo flattening from an ensemble of projected stream tracks. In that approach, each stream is forward modelled with a particle spray generator and fit within a Bayesian inference pipeline to obtain an individual posterior on the halo flattening while explicitly addressing projection driven degeneracies and reducing severe multimodality through physically motivated prior restrictions. Although any single stream typically yields only weak constraints, the resulting set of individual posteriors can be coherently combined through hierarchical reweighting \citep{Hogg2010} to infer hyperparameters describing the population distribution of halo flattening across the full sample. Tests on mock catalogues demonstrated that, for sample sizes comparable to \textsc{STRRINGS}, the method recovers informative population level constraints and cleanly distinguishes between oblate, spherical, and prolate halo populations even when individual posteriors remain broad. The population inference is computationally scalable with overall cost growing approximately linearly with the number of streams, making it well matched to the increasing stream samples expected from upcoming wide-field surveys.

In this paper we take the next step by applying the population level stream modelling framework developed in \citetalias{Chemaly2026} to the \textsc{STRRINGS} catalogue (\citetalias{Sola2025}). We forward model 32 streams within an axisymmetric halo potential and infer posterior constraints on the halo flattening for each host galaxy. We then combine these individual posteriors in a hierarchical model to recover the underlying population distribution of halo flattening for nearby galaxies hosting dynamically simple minor merger streams. Finally, we place the observationally inferred population in context by comparing it to expectations from cosmological simulations \citep[e.g.][]{Chua2019,Prada2019}. This work yields a photometry only measurements of the population distribution of dark matter halo flattening from stellar streams beyond the Local Group.

The paper is organised as follows. Section~\ref{sec:data} summarises the data used in this work. Section~\ref{sec:methods} describes the stream model and parameters used to fit both individual streams and population parameters. Section~\ref{sec:results} presents the 32 individual stream fits and combines them to constrain the population level distribution of halo flattening. Section~\ref{sec:discussion} discusses limitations, and Section~\ref{sec:conclusions} concludes with a summary of results and implications for future work.

\section{Data}\label{sec:data}

We use the \textsc{STRRINGS} (\textit{STReams in Residual Images of Nearby GalaxieS}) catalogue of extragalactic stellar streams presented in \citetalias{Sola2025}. \textsc{STRRINGS} was constructed from the DESI Legacy Imaging Surveys \citep{Dey2019}, leveraging the fact LSB tidal features are often most easily detected in residual images, where the smooth light distribution of bright sources has been modelled and subtracted. In this context, residual maps suppress the high surface brightness contribution from the main galaxy (and other luminous contaminants), enhancing faint asymmetries and extended structures and enabling a systematic search for narrow, elongated streams in nearby systems.

The \textsc{STRRINGS} selection was performed by visually inspecting residual images to identify thin, coherent, stream-like features consistent with minor merger debris and suitable for dynamical modelling (\citetalias{Sola2025}). For each candidate, the stream was then manually segmented using \texttt{Jafar} \citep{Sola2022} to isolate the pixels associated with the feature and to define the stream track, a one dimensional ridge line that traces the mean path of the stream on the sky. In \textsc{STRRINGS}, this ridge line is encoded as $r(\phi)$: the projected radial distance, r, from the host galaxy centre as a function of (unwrapped) polar angle $\phi$, measured in discrete angular bins (\citetalias{Sola2025}). Operationally, the track is extracted by sweeping through angular bins and, within each bin, collapsing the segmented residual light into a one dimensional radial profile. A one dimensional Gaussian model is then fitted to the mean residual brightness as a function of radius in that angular slice. The Gaussian mean provides an estimate of the stream radial distance (i.e.\ the track value $r$ at that $\phi$), the standard deviation encodes the stream width in projection, and the amplitude is proportional to the stream flux contribution within the bin. In this work, we restrict the analysis to the stream track by fitting the mean radial distance in fixed angular bins because our stream modelling does not accurately reproduce the stream width nor flux distribution (see Section~\ref{subsec:stream_model} for further details).

From the 35 streams in the \textsc{STRRINGS} catalogue, we retain 32 for analysis. We exclude NGC1084, NGC1121, and PGC021008. In the latter case, the stream is sufficiently bright that it was explicitly modelled during the host galaxy light subtraction and therefore does not appear cleanly in the residual images used in our pipeline. For NGC1084, our forward modelling failed to converge to a satisfactory fit; upon closer inspection, the features might more closely resembles a grand design spiral than a genuine tidal stream, which likely explains the failure of the model. NGC1121 is a confirmed, prominent three loop stream. Owing to its exceptional angular extent and morphological complexity, it requires tailored modelling beyond the scope of this automated analysis. We refer to Erkal et al. (in prep.) for a dedicated study and complete fit of NGC1121. The remaining 32 streams were successfully and automatically fitted as part of this work.

Unless stated otherwise, all distances are expressed in physical units using the host galaxy distance, average of 60~Mpc, adopted by the original \textsc{STRRINGS} paper. For full details of the imaging, segmentation, and track extraction methodology, we refer the reader to \cite{Sola2025}.

\section{Methods}\label{sec:methods}

Our goal is to infer the flattening of the dark matter halo hosting each stream. Since cosmological simulations show that dark matter halos tend to be quite axisymmetric \citep{Prada2019}, we model the host galaxy potential as an axisymmetric Navarro–Frenk–White (NFW) halo whose shape is controlled by a single flattening parameter. Given a set of parameters, we generate a mock stellar stream using a particle spray forward model and compare its projected track to the observed stream track through a likelihood function.

This pipeline was introduced in \citetalias{Chemaly2026}, where it was validated on mock data. The inference proceeds in two stages. First, for each stream, we perform Bayesian forward modelling to obtain a posterior distribution over the halo flattening. Second, we combine the individual posteriors in a hierarchical Bayesian framework to infer the population distribution of halo flattening across the sample.

In the present work we apply it to observational data and implement several modifications to improve robustness to observational systematics and to mismatches between the idealised model and the underlying gravitational potential.

\subsection{Stream modelling}\label{subsec:stream_model}

As in \citetalias{Chemaly2026}, we generate streams using the Lagrangian particle spray method \citep[e.g.][]{Johnston1998,Kuepper2012,Gibbons2014,Fardal2015}, integrating test particles in the combined potential of the host and satellite with a leapfrog integrator \citep{Hockney1988}. The spray prescription approximates tidal stripping by releasing particles from the Lagrangian points of the progenitor over time, producing a dynamically cold debris structure whose projected track can be compared directly to the observed track.

In \citetalias{Chemaly2026}, we obtained the ordering along the stream by unwrapping the orbit of every particle individually. From there, by taking into account the birth angle and the final unwrapped angle, we were able to order every particle from trailing to leading. For real systems, this can fail in edge cases, in particular when the projected orbit is not strictly monotonic in polar angle, leading to ambiguous ordering and discontinuities in the unwrapped representation. To address this, we adopt the ordering procedure of \cite{Gibbons2014} (see their sub-section 3.4), which produces a more stable mapping of the simulated debris onto a one-dimensional coordinate along the stream. In practice, this improves robustness for streams with turning points or locally complex projected geometry.

Following \citetalias{Chemaly2026}, we fix one of the on-sky projected coordinate of the progenitor when fitting each stream. In a subset of the streams in \textsc{STRRINGS}, a progenitor candidate is visible, allowing this coordinate to be fixed at the observed location. In most of the streams sample, however, no progenitor is confidently identified. In these cases, we place the progenitor at the highest surface brightness region near the centre of the track, which typically corresponds to the densest region of the stream track (\citetalias{Sola2025}). We note that future work will relax this assumption by sampling the progenitor's position along the track.

We additionally introduce a simple time-dependent progenitor mass loss model. When no progenitor is visible, the progenitor mass is linearly reduced to zero over the integration interval, effectively suppressing the pronounced kink (or S-shape) near the remnant that is commonly present in idealised simulations but is unresolved in our extragalactic survey. When a progenitor candidate is visible, we instead hold the progenitor mass fixed. Empirically, this modification has little impact on the large scale track morphology, but improves stability of the fit around the (often uncertain) progenitor region.

Unlike mock experiments, where the same forward model generates both the synthetic data and the prediction so that the two are guaranteed to be sampled consistently, real-data inference can be biased if the number of simulated particles is too small. In that case, parameter combinations may be disfavoured not because they provide a poor match, but because the modelled stream does not contain enough samples. To mitigate this, we deliberately oversample each forward model. We do this by setting the total number of particles to the highest between $N_{\rm tot}=10^4$ or $1,500$ particles per data points along the track. This condition was empirically tested to allow for a good fit of the projected track while respecting the imposed condition on the model versus data uncertainty discussed in the Sub-Section \ref{subsec:individual_likelihood}. Since we are leveraging \href{https://github.com/David-Chemaly/StreaMAX}{\texttt{StreaMAX} \textsc{\Large{\scalebox{0.8}{\faGithub}}}} (\citetalias{Chemaly2026}), the increased particle counts remain computationally tractable, and individual fits typically complete within a day on a 44-core CPU. 

\begin{figure*}
    \centering

    \begin{minipage}{0.48\textwidth}
        \centering
        \includegraphics[width=\linewidth]{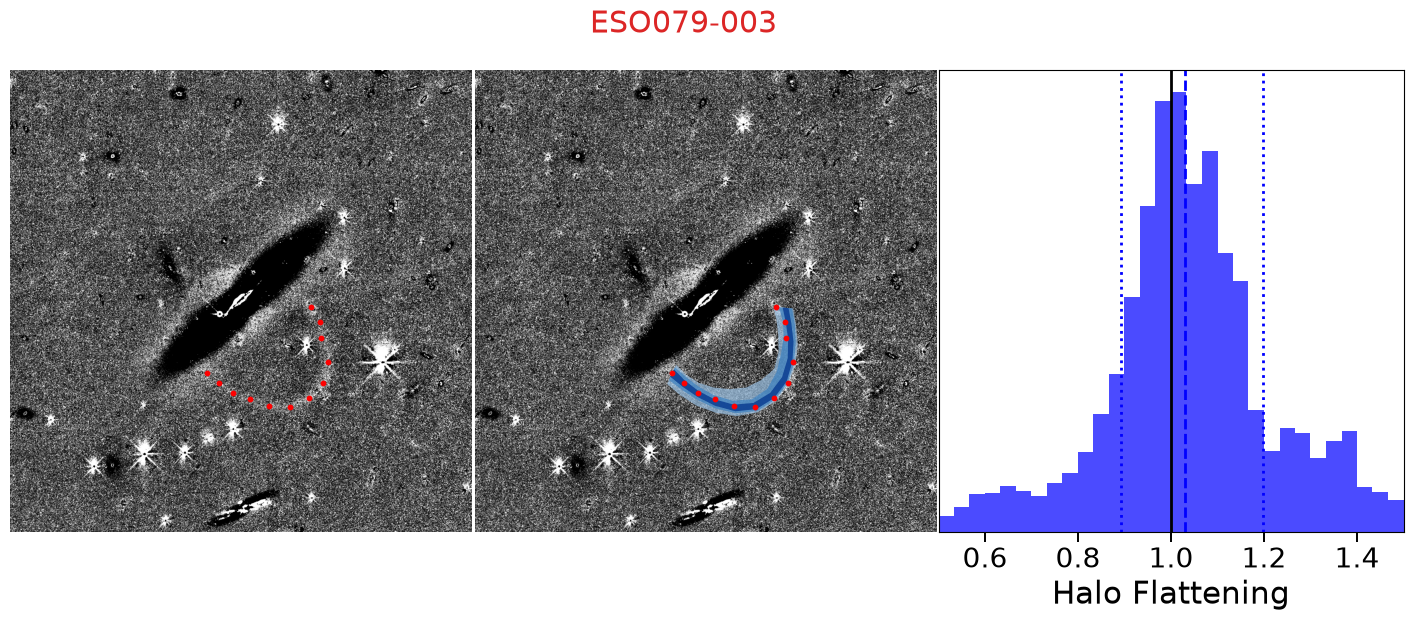}
    \end{minipage}
    \begin{minipage}{0.48\textwidth}
        \centering
        \includegraphics[width=\linewidth]{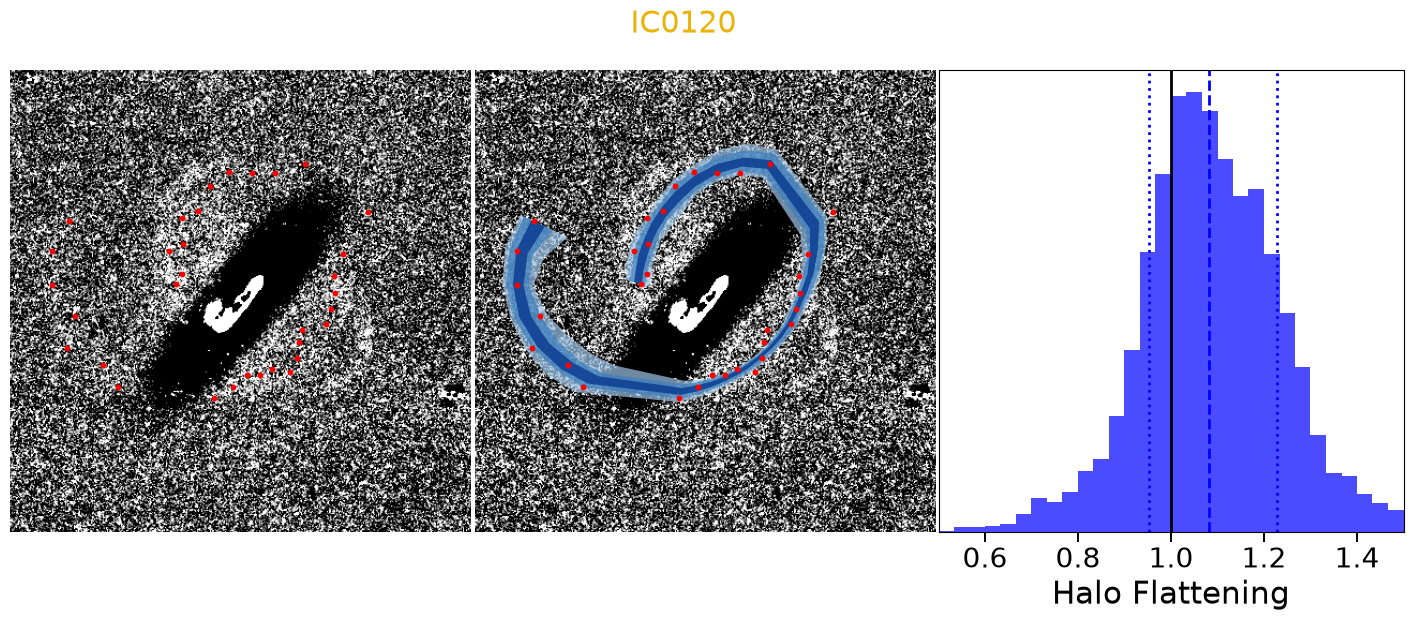}
    \end{minipage}

    \begin{minipage}{0.48\textwidth}
        \centering
        \includegraphics[width=\linewidth]{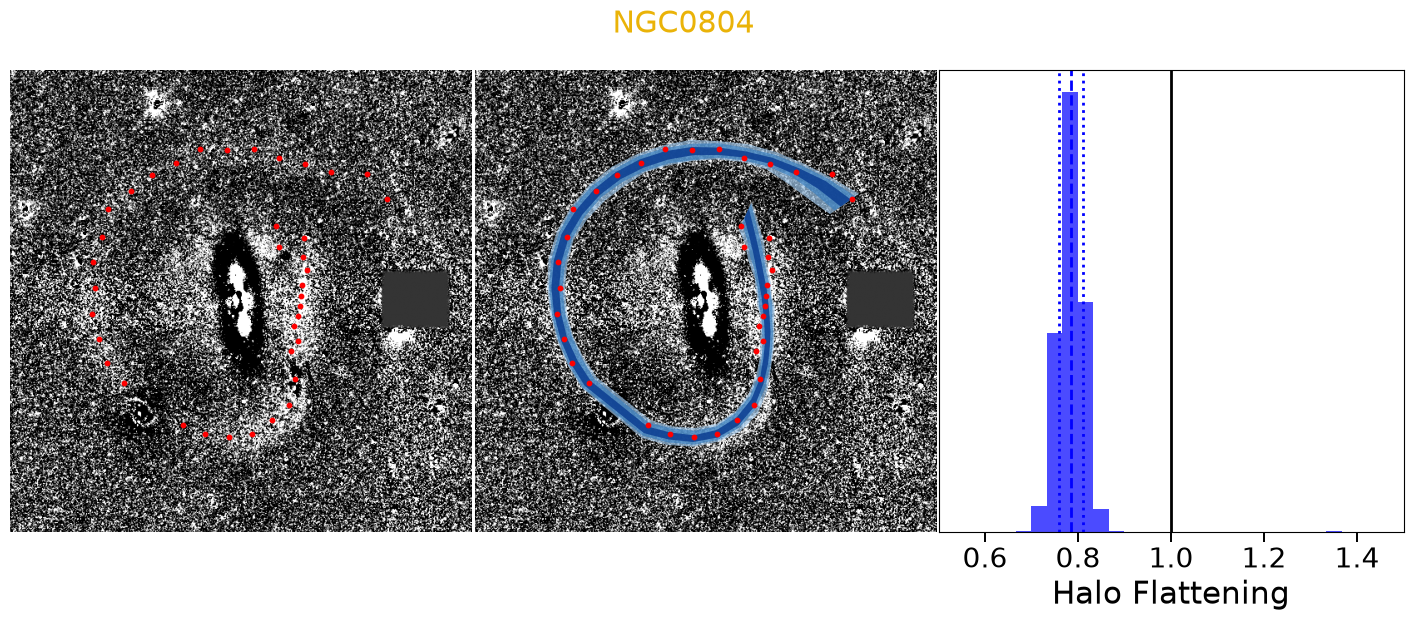}
    \end{minipage}
    \begin{minipage}{0.48\textwidth}
        \centering
        \includegraphics[width=\linewidth]{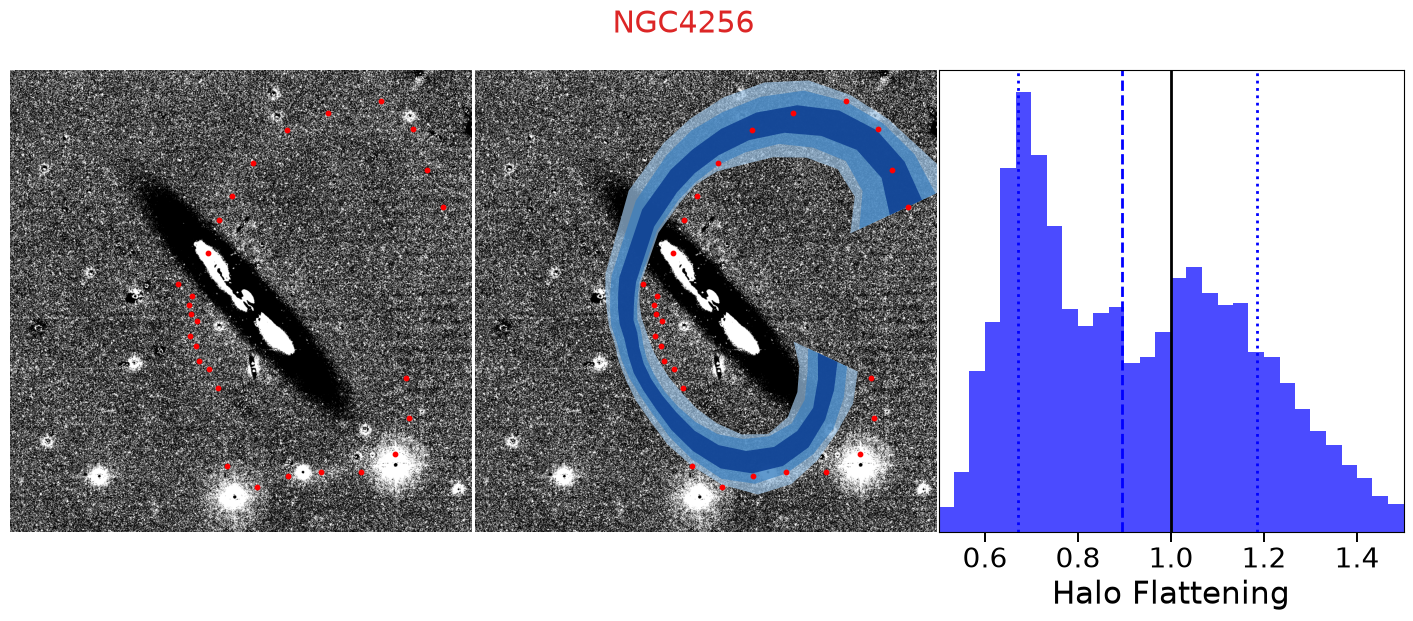}
    \end{minipage}

    \begin{minipage}{0.48\textwidth}
        \centering
        \includegraphics[width=\linewidth]{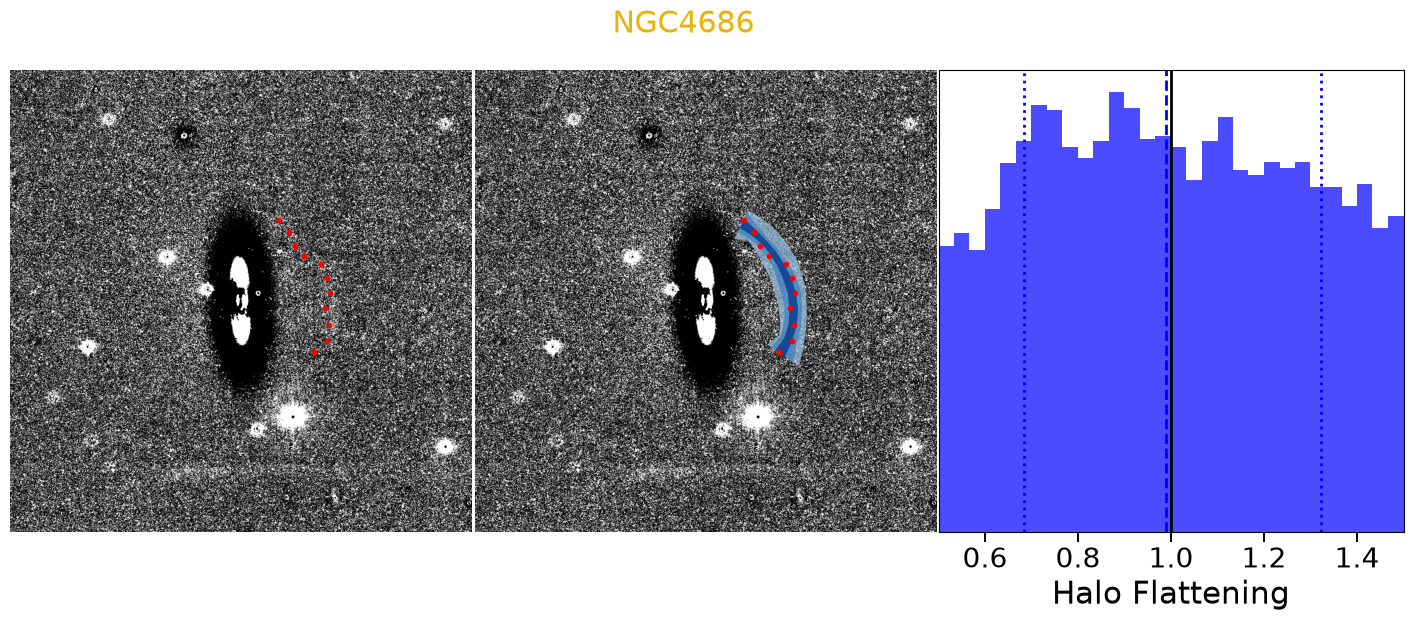}
    \end{minipage}
    \begin{minipage}{0.48\textwidth}
        \centering
        \includegraphics[width=\linewidth]{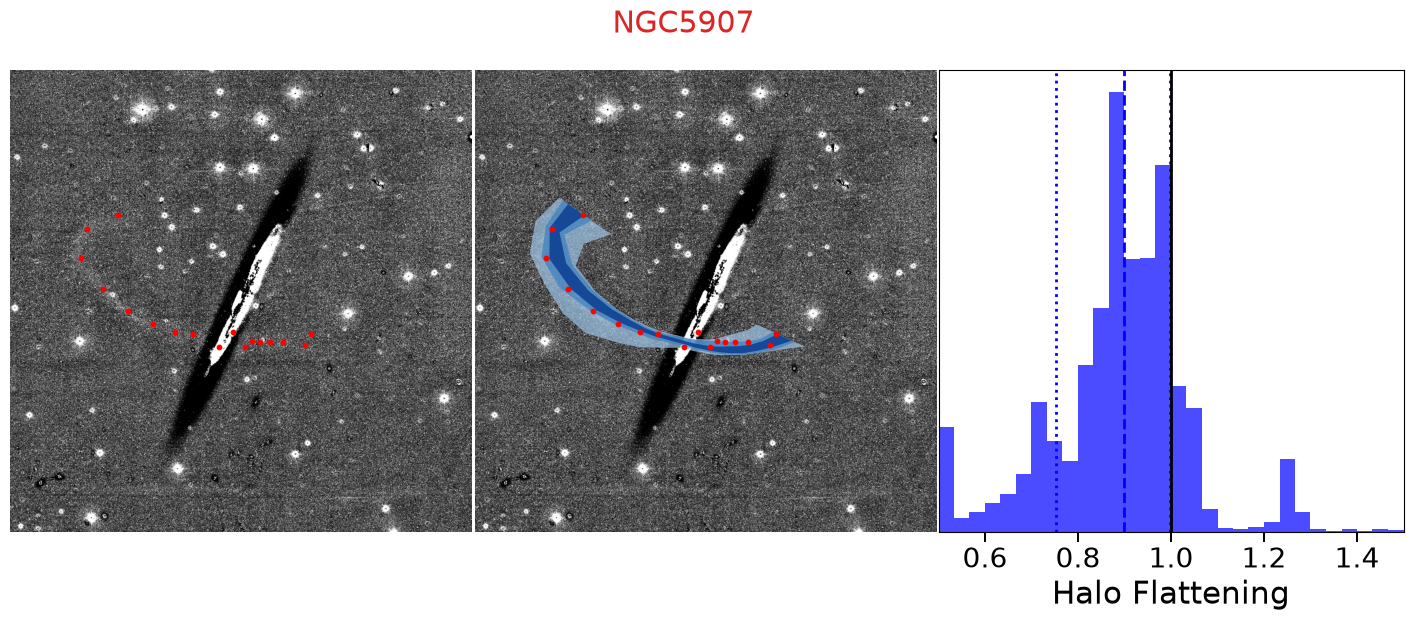}
    \end{minipage}

    \begin{minipage}{0.48\textwidth}
        \centering
        \includegraphics[width=\linewidth]{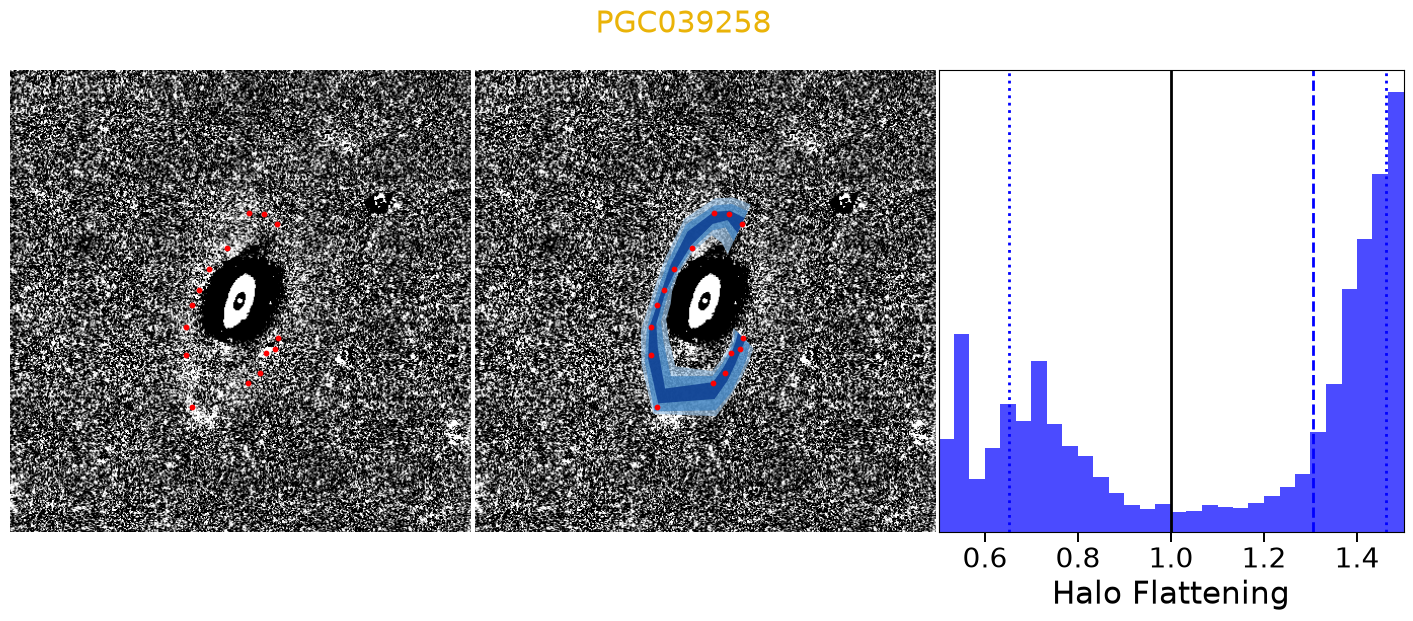}
    \end{minipage}
    \begin{minipage}{0.48\textwidth}
        \centering
        \includegraphics[width=\linewidth]{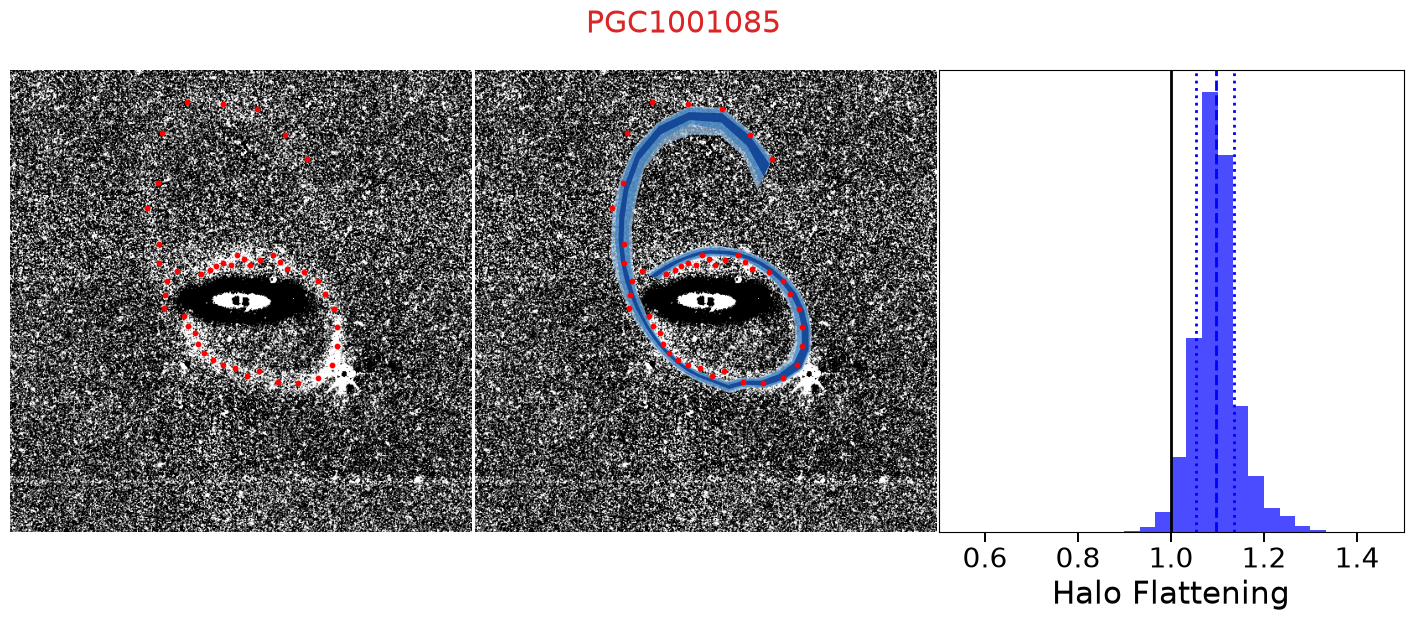}
    \end{minipage}

    \caption{Examples of fits for eight streams chosen from the STRRINGS catalogue. The residual image with the extracted stream track (red) is shown in the left panel. The middle panel overlays the best fit, and the right panel shows the posterior distribution of the dark matter halo flattening assuming an axisymmetric NFW potential. The black vertical line marks the spherical case ($q=1$), with oblate configurations ($q<1$) to the left and prolate configurations ($q>1$) to the right. The dashed blue line indicates the posterior median, while the dotted blue lines show the 16th and 84th percentiles. The remaining streams in STRRINGS are shown in Appendix~\ref{app:B}. In total, we performed 32 fits. The colour of each stream name indicates its subsample classification, introduced in Subsection~\ref{sec:population}.}
    \label{fig:example_individual}
\end{figure*}

\subsection{Model parameters}

Our baseline model follows \citetalias{Chemaly2026}. The host halo is represented by an axisymmetric NFW potential, parameterised by a characteristic mass and scale radius, together with three extra parameters controlling the halo orientation and flattening. We adopt the same orientation and flattening parameterisation as \citetalias{Chemaly2026}: we sample a three component vector $(x,y,z)$ that defines the orientation of the axis of symmetry, while the vector norm maps to the halo flattening. The disrupting progenitor is modelled with a Plummer potential, parameterised by a mass and scale radius. Setting x and y as the on-sky coordinates and z along the line of sight, the present day phase space coordinates of the progenitor are specified by $(x_f,z_f,v_{x_f},v_{y_f},v_{z_f})$, with the projected $y_f$ coordinate held fixed to reflect our assumed knowledge of the progenitor's approximate location in the sky plane (Section~\ref{subsec:stream_model}). Finally, the integration time sets the disruption timescale over which particles are released and evolved. The priors for these dynamical parameters are unchanged from \citetalias{Chemaly2026}. For convenience, we refer to the schematic of the coordinate system in \citetalias{Chemaly2026} (their Figure~1) and summarise all parameters and priors in the Appendix~\ref{app:A}.

When fitting mock data, \citetalias{Chemaly2026} could correctly assume that the model used for inference matched the model used to generate the mock data. For real streams, this is not the case: our axisymmetric halo model neglects baryonic components and other sources of non-axisymmetry, and therefore cannot be expected to reproduce every track perfectly. To account for this, we include an additional, strictly positive scatter term to account for unmodeled systematics \citep{Koposov2023}, $\sigma_{\rm sys}$, in the individual likelihood (see Section~\ref{subsec:individual_likelihood}). Conceptually, $\sigma_{\rm sys}$ inflates the effective uncertainty on the track and absorbs residual mismatch between the model and the data. The parameter is self regulated by the likelihood: small values indicate that the axisymmetric model can reproduce the track within the reported measurement errors, while larger values indicate that additional unmodelled physics (or systematics) are required to explain the residuals. We adopt a uniform prior $\sigma_{\rm sys}\sim \mathcal{U}(0,25)\,\mathrm{kpc}$. The prior is purposely chosen to be very wide, guaranteeing that the self regulation of this extra systematic term is not affected by the limits. The prior range is intentionally broad to avoid artificially constraining this term: a value of $25\,\mathrm{kpc}$ corresponds to an extremely large track uncertainty, far exceeding both the typical measurement errors and the projected stream widths in STRRINGS. In practice, the posterior values of $\sigma_{\rm sys}$ are always much smaller, indicating that the inference is effectively driven by the data rather than by the prior limits. With this extra parameter, our model has a total of 14 parameters (see Appendix~\ref{app:A}).

Our parameter of interest is the halo flattening, $q$, inferred for each stream. All remaining parameters are treated as nuisance and marginalised over. We also report the inferred $\sigma_{\rm sys}$ as a diagnostic of model adequacy and to ensure that the morphological constraints are not driven by unmodelled systematics.

\subsection{Individual likelihood}\label{subsec:individual_likelihood}

We first fit each stream independently using Bayesian inference. Given data $D$ and a model prediction $M(\theta)$ parametrised by $\theta$, the posterior distribution of the parameters is

\begin{equation}\label{eq:bayes}
  P(\theta \mid D)
  \propto P(D \mid \theta)\,\pi(\theta),
\end{equation}
where $\pi(\theta)$ is the prior and $P(D\mid\theta)$ is the likelihood of the data given the model prediction $M(\theta)$. Note that the proportionality is due to the fact that we ignore the evidence since we do not compare different models.

The observational data consist of $N$ radial distances of the tracks measured in fixed angular bins,
$r_{\mathrm{data}_i} \equiv r_{\mathrm{data}}(\phi_i)$, with associated uncertainties
$\sigma_{\mathrm{data}_i}$ provided by \textsc{STRRINGS}. 
Following the assumptions of \textsc{STRRINGS}, we adopt Gaussian measurement errors:

\begin{equation}
  r_{\mathrm{data}_i} \sim 
  \mathcal{N}\!\left(
    r_{\mathrm{model}_i}(\theta),
    \sigma_{\mathrm{data}_i}^2
  \right),
  \qquad i=1,\dots,N.
\end{equation}

Since we are binning the data, the likelihood is independent of the angle and only a function of the one dimensional stream track (i.e. the radial distance). Working with the logarithm for numerical stability, we can immediately write the log-likelihood as

\begin{equation}
\begin{aligned}
\log P(D\mid\theta)
= -\frac{1}{2}\sum_{i=1}^{N}
\Bigg[
    \frac{\big(r_{\mathrm{model}_i}(\theta)-r_{\mathrm{data}_i}\big)^2}
         {\sigma_{\mathrm{data}_i}^{2}}
     \\
    \;+\; \log\!\Big(2\pi\sigma_{\mathrm{data}_i}^{2})
\Bigg].
\end{aligned}
\label{eq:loglike_base}
\end{equation}

This is the exact framework set in \citetalias{Chemaly2026} (see their Equation 16). Comparatively, we introduce an additional parameter $\sigma_{\mathrm{sys}}$ that accounts for systematics between the idealised model and the true gravitational potential. That leads to the likelihood used in this work:

\begin{equation}
\begin{aligned}
\log P(D\mid\theta)
= -\frac{1}{2}\sum_{i=1}^{N}
\Bigg[
    \frac{\big(r_{\mathrm{model}_i}(\theta)-r_{\mathrm{data}_i}\big)^2}
         {\sigma_{\mathrm{data}_i}^{2}+\sigma_{\mathrm{sys}}^{2}}
     \\
    \;+\; \log\!\Big(2\pi\big(\sigma_{\mathrm{data}_i}^{2}+\sigma_{\mathrm{sys}}^{2}\big)\Big)
\Bigg].
\end{aligned}
\label{eq:loglike_sys}
\end{equation}

Equivalently, this corresponds to inflating the measurement model to
\begin{equation}
  r_{\mathrm{data}_i} \sim
  \mathcal{N}\!\left(
    r_{\mathrm{model}_i}(\theta),
    \sigma_{\mathrm{data}_i}^2 + \sigma_{\mathrm{sys}}^2
  \right),
  \qquad i=1,\dots,N.
\end{equation}
Because $\sigma_{\mathrm{sys}}$ enters both the residual and the normalisation term, the likelihood converges to the minimal additional variance required by the data.

Before evaluating the likelihood we enforce two conditions ensuring that the forward model provides a reliable prediction. First, the modelled stream must cover the angular range of the data: each observational bin must contain at least three model particles. Otherwise, the likelihood is set to a very bad but still finite value. Second, the uncertainty on the modelled stream track must be nine times smaller than the observational uncertainty. We choose nine because it is roughly an order of magnitude and has a clean square root. In other words, this allows to say that the standard deviation of the model has to be three times smaller than the standard deviation of the data. For each bin we estimate the model uncertainty

\[
\sigma_{\rm model} \simeq \frac{\sigma_r}{\sqrt{n}},
\]

where $\sigma_r$ is the radial standard deviation of particles in the bin and $n$ the number of particles. We require

\begin{equation}
\sigma_{\rm model}^2 \le \frac{\sigma_{\rm data}^2}{9}.
\end{equation}

This ensures that the likelihood function is smooth and is not affected by noise caused by the random sampling of streams by particles. Parameter proposals violating this condition are assigned a very low likelihood value, low enough to be completely rejected but not set to $-\infty$ to aid sampler convergence.

As in \citetalias{Chemaly2026}, we explore the posterior using dynamic nested sampling implemented in \texttt{dynesty} \citep{Higson2019,Speagle2020,Koposov2022}, starting from 2000 live points and using the \textit{rslice} sampler. This approach efficiently handles the high dimensionality, degeneracies, and multimodality of the posterior distributions.

\subsection{Population likelihood}\label{subsec:population_likelihood}

Beyond individual stream fits, our main objective is to constrain the population distribution of halo flattening. Individual streams typically yield broad and sometimes multimodal posteriors on the flattening, but these can be combined across the ensemble of streams around galaxies using hierarchical Bayesian inference to obtain robust constraints on population level parameters and compare them against expectations from cosmological simulations \citep[e.g.][]{Chua2019,Prada2019}.

We follow the hierarchical importance sampling framework derived in \citetalias{Chemaly2026}. For each stream $N$, we denote the data by $d_n$ and the full set of model parameters by $\theta_n$. We introduce hyperparameters $\alpha$ describing the population model for the halo flattening $q$ with hyperprior $\pi(\alpha)$. Assuming conditional independence of the data given the individual parameters and that $\theta_n$ are drawn from a population prior $\pi(\theta\mid\alpha)$, the hierarchical posterior is

\begin{equation}
    P(\alpha\mid\{d_n\}) \propto \prod_{n=1}^{N}
    \bigg[\int L(d_n\mid\theta_n)\,\pi(\theta_n\mid\alpha){\rm d}\theta_n \bigg] \pi(\alpha).
\label{eq:raw_hierarchical}
\end{equation}

Evaluating Eq.~\ref{eq:raw_hierarchical} directly would require refitting each stream for every proposed $\alpha$ which is computationally unfeasible. Instead, we reuse posterior samples from the individual fits and compute $L(d_n\mid\alpha)$ via importance reweighting \citep{Hogg2010}.

Concretely, we first perform independent Bayesian fits for each stream using an interim prior $\pi(\theta_n)$, which in our case includes a uniform prior on flattening $q\in[0.5,1.5]$. Rewriting Bayes' theorem (see Eq.~\ref{eq:bayes}), we get 

\begin{equation}
    L(d_n\mid\theta_n)\ \propto\ \frac{P(\theta_n\mid d_n)}{\pi(\theta_n)}.
\label{eq:posterior_to_likelihood}
\end{equation}
Substituting Eq.~\ref{eq:posterior_to_likelihood} into Eq.~\ref{eq:raw_hierarchical} yields
\begin{equation}
    P(\alpha\mid\{d_n\}) \propto  \prod_{n=1}^{N} \bigg[
    \int P(\theta_n\mid d_n)\,\frac{\pi(\theta_n\mid\alpha)}{\pi(\theta_n)}\,{\rm d}\theta_n \bigg] \pi(\alpha).
\label{eq:hier_reweight}
\end{equation}

The integral in Eq.~\eqref{eq:hier_reweight} is an expectation over the known posterior $P(\theta_n\mid d_n)$ and can be approximated using posterior samples obtained from the individual fits. Let $\{\theta_{n,i}\}_{i=1}^{K_n}$ be $K_n$ samples from  $P(\theta\mid d_n)$, we get
\begin{equation}
P(\alpha\mid\{d_n\}) \propto \prod_{n=1}^{N}
\left[
\frac{1}{K_n}\sum_{i=1}^{K_n}
\frac{\pi(\theta_{n,i}\mid\alpha)}{\pi(\theta_{n,i})}
\right]\pi(\alpha).
\label{eq:hier_mc}
\end{equation}
Because the samples $\{\theta_{n,i}\}$ are already drawn from the posterior $P(\theta_n\mid d_n)$, the posterior density itself does not appear in the sum: the expectation reduces to the sample mean of the importance weight $\pi(\theta_{n,i}\mid\alpha)/\pi(\theta_{n,i})$.

We decompose $\theta=(q,\psi)$, where $q$ is the halo flattening and $\psi$ collects nuisance parameters. Since $\alpha$ only affects the prior over $q$ while the priors on $\psi$ remain fix in both stages and the prior on $q$ is always uniform, the ratio can be simplified:
\begin{equation}
\frac{\pi(\theta\mid\alpha)}{\pi(\theta)}=\frac{\pi(\psi)\pi(q\mid\alpha)}{\pi(\psi)\pi(\theta)}=\frac{\pi(q\mid\alpha)}{\pi(q)}\propto\pi(q\mid\alpha).
\end{equation}
The working expression used throughout this paper is therefore
\begin{equation}
P(\alpha\mid\{d_n\}) \propto \prod_{n=1}^{N}
\left[
\frac{1}{K_n}\sum_{i=1}^{K_n}
\pi(q_{n,i}\mid\alpha)
\right]\pi(\alpha).
\label{eq:hier_final}
\end{equation}

This approach reuses the individual posterior samples and yields a second-stage inference instead of the initial $\mathcal{O}(N^2)$ from Eq.~\ref{eq:raw_hierarchical} making it well suited to the large samples of external streams expected from \textit{Euclid} and Rubin/LSST. A key requirement is that the individual posteriors have enough samples to support the high probability regions defined by the population prior $\pi(q\mid\alpha)$ since no new per stream samples are generated in the reweighting step. In our case, the individual posteriors are typically broad and cover most of the interim prior $\pi(q)$. Therefore, we simply enforce the population prior to always be within the range of the interim prior for any $\alpha$.

We adopt the same population model as in \citetalias{Chemaly2026}: a Gaussian distribution in $q$ with mean $\mu_{\rm pop}$ and scatter $\sigma_{\rm pop}$, truncated to $q\in[0.5,1.5]$. We set $\alpha=(\mu_{\rm pop},\sigma_{\rm pop})$ and explore $P(\alpha\mid\{d_n\})$ using dynamic nested sampling with \texttt{dynesty}, starting from 500 live points. For full derivations, implementation details, and validation on mock catalogues, we refer the reader to \citetalias{Chemaly2026}.

\section{Results}\label{sec:results}

\subsection{Individual fit}\label{sec:individual}

We first fit each stream track independently using Bayesian inference, and obtain posterior constraints on the dark matter halo flattening assuming an axisymmetric NFW model. We perform individual fits for 32 streams in \textsc{STRRINGS} out of 35. Figure~\ref{fig:example_individual} illustrates the results for eight systems: the left panel shows the residual image with the \textsc{STRRINGS} track overlaid, the middle panel compares the observed track to the best fitting forward model in projection, and the right panel shows the resulting marginal posterior on the halo flattening. The corresponding figures for the remaining 24 streams are presented in Appendix~\ref{app:B}, and as an example of a complete fit the full joint posterior of all model parameters for NGC0804 is shown in Appendix~\ref{app:corner}. The colour of each name represents the subsample classification which will be introduced in Subsection~\ref{sec:population}.

Overall, the best fitting projected tracks reproduce the data well across the sample, indicating that the forward model can capture the dominant geometrical features of the observed streams. However, the informative power of individual streams varies substantially: in some cases the posterior on flattening remains broad and close to the uniform prior (i.e. NGC4686), while in others the data yield noticeably tighter constraints (i.e. NGC5263, NGC5907, PGC1001085). In Section~\ref{subsec:properties} we quantify how this constraining power correlates with stream properties, namely angular extent, physical length, mean radius and radial spread.

Consistent with the mock data experiments in \citetalias{Chemaly2026}, some systems exhibit pronounced multimodality between oblate and prolate solutions (e.g. PGC039258), reflecting projection driven degeneracies whereby distinct three dimensional potentials can produce similar projected tracks (see Fig.~6 of \citetalias{Chemaly2026}). Nevertheless, the projected stream geometry still retains information about the underlying three dimensional potential \citep[see also][]{Nibauer2025a}, such that these modes are not always symmetric and one solution is often preferred.

\begin{figure}
    \centering
    \includegraphics[width=\linewidth]{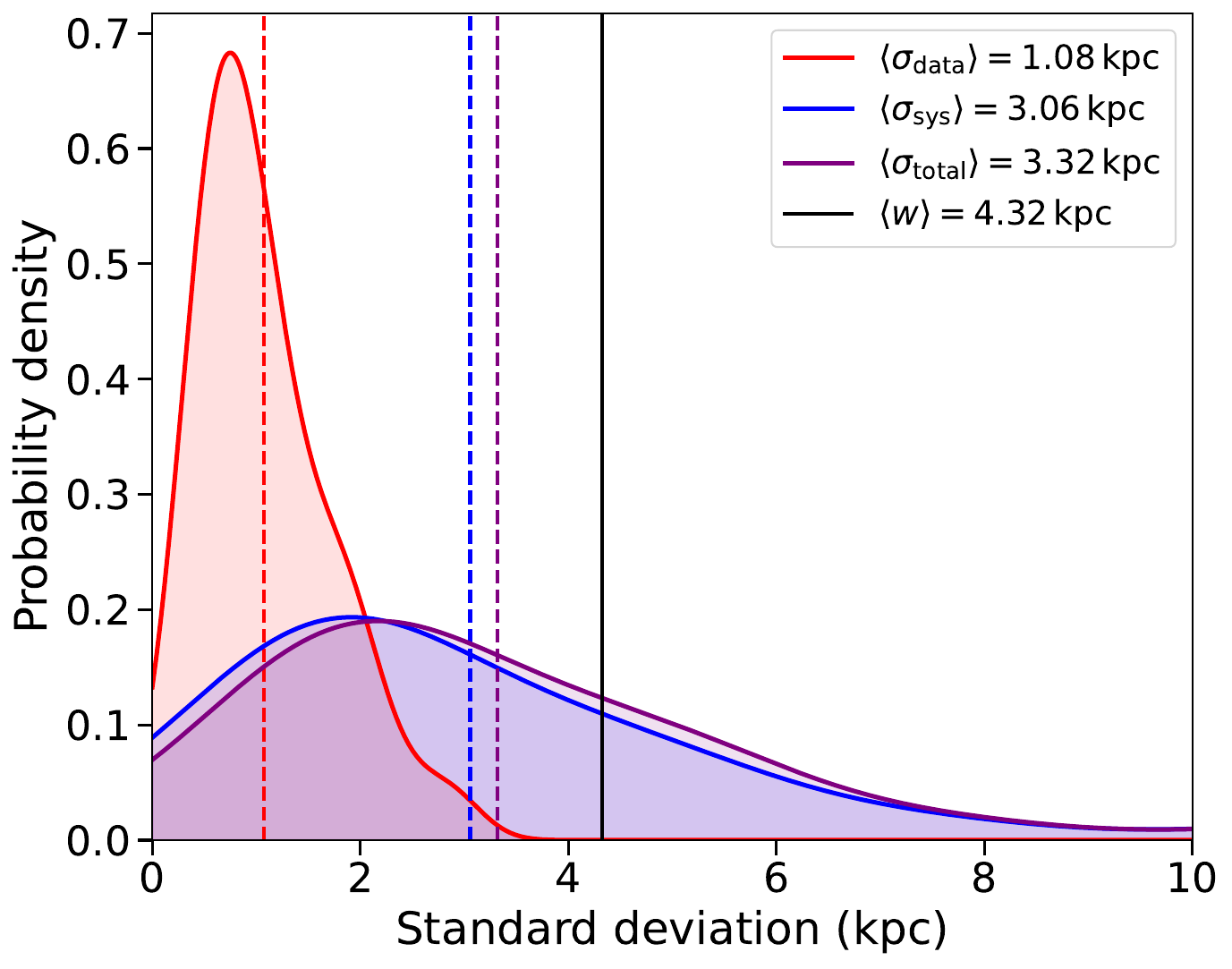}
    \caption{Distribution of the per stream track uncertainty $\sigma_{\rm data}$ (red), the inferred additional scatter term $\sigma_{\rm sys}$ (blue), and the total effective scatter $\sigma_{\rm total}=\sqrt{\sigma_{\rm data}^2+\sigma_{\rm sys}^2}$ (purple). Each distribution is constructed from one summary value per stream ($N=32$) and smoothed with a kernel density estimate (KDE). The mean of each distribution is shown in the legend, and the black vertical line represents the mean projected width of the stream in STRRINGS.}
    \label{fig:sigma}
\end{figure}

\begin{figure*}
    \centering
    \includegraphics[width=\linewidth]{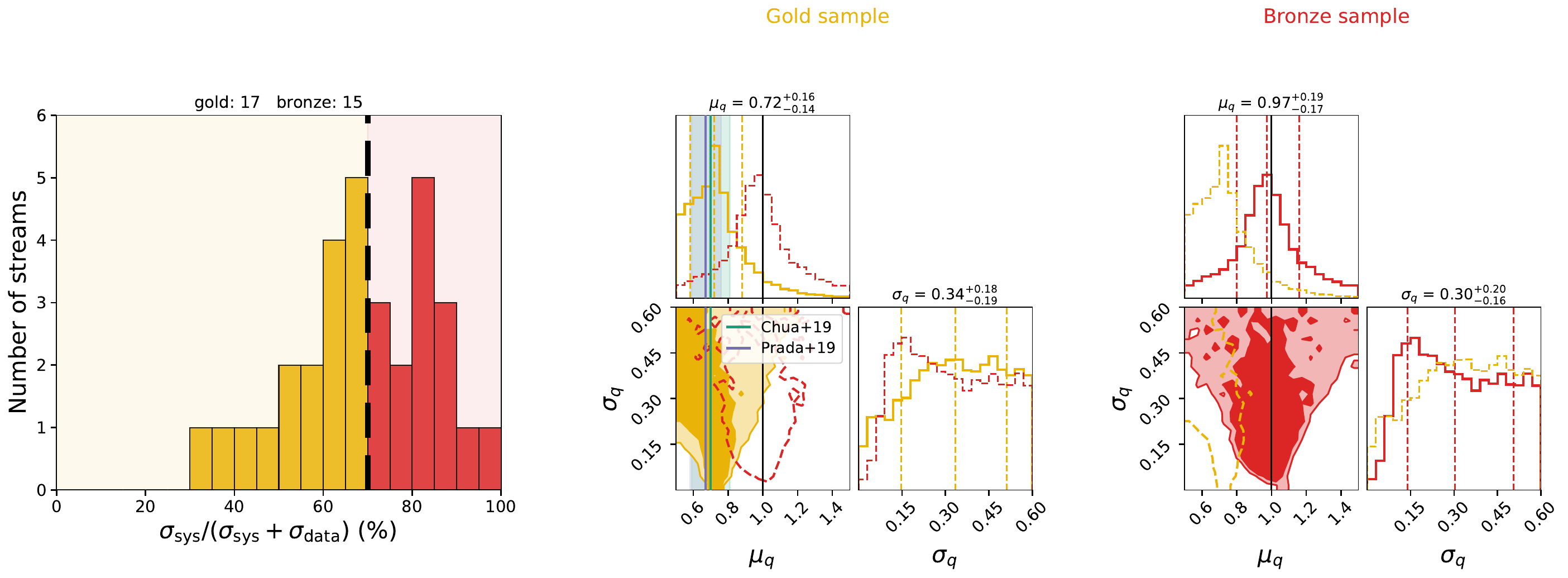}
    \caption{Posterior distributions of the population hyperparameters $(\mu_q,\sigma_q)$ obtained by splitting the sample according to the fractional contribution of the systematic term to the total uncertainty, $\sigma_{\rm sys}/(\sigma_{\rm sys}+\sigma_{\rm data})$ (left panel). Streams with a ratio below 70\% are classified as the \textit{gold} sample (middle; gold), while those above 70\% form the \textit{bronze} sample (right; red). Each corner overlays the other sample's $1\sigma$ contour (dashed) for comparison. The population inference assumes a truncated Gaussian model for the halo flattening distribution. The black vertical line marks the spherical case ($\mu_q=1$), with oblate configurations ($\mu_q<1$) to the left and prolate configurations ($\mu_q>1$) to the right. The dashed lines, coloured by sample, indicate in order the 16th, 50th and 84th percentiles of the posterior distributions. In the gold panel, the vertical bands (shown on both the $\mu_q$ marginal and the joint posterior) mark the $1\sigma$ inner-halo flattening predicted by baryonic cosmological simulations, for comparison with our measurement: $q=0.70\pm0.11$ (\citealt{Chua2019}, IllustrisTNG) and $q=0.67\pm0.09$ (\citealt{Prada2019}, Auriga).
    }
    \label{fig:population}
\end{figure*}

Figure~\ref{fig:sigma} summarises the posterior behaviour of the additional scatter term $\sigma_{\rm sys}$ introduced in Eq.~\ref{eq:loglike_sys} for all 32 streams. For each stream, $\sigma_{\rm data}$ is taken from \textsc{STRRINGS} as the per bin radial uncertainty of the extracted track, and we compress it to a single representative value by averaging over angular bins. We do so to allow for a simpler comparison to other parameters. For $\sigma_{\rm sys}$, we use the posterior median of $\sigma_{\rm sys}$ from the individual fit as a summary statistic. This yields one value per stream, from which we form the distributions shown in Fig.~\ref{fig:sigma}, where each distribution is smoothed with a kernel density estimate (KDE) for clarity.

The inferred $\sigma_{\rm sys}$ is consistently larger than the nominal track uncertainties, with $\langle\sigma_{\rm sys}\rangle \simeq 3.12\,{\rm kpc}$ compared to $\langle\sigma_{\rm data}\rangle \simeq 1.08\,{\rm kpc}$, implying a typical effective scatter $\langle\sigma_{\rm total}\rangle \simeq 3.37\,{\rm kpc}$. In absolute terms this scale is modest: it is small compared to typical stream radii (around $5\%$) and is of the same order as (but still below) the typical projected stream width ($\langle w\rangle \simeq4.32\,{\rm kpc}$ in STRRINGS), indicating that the forward model can reproduce the observed tracks to within a tolerance smaller than the projected stream width. Some likely drivers of this mismatch could be the (i) track extraction from residual images and (ii) binning in angle. The \textsc{STRRINGS} track is obtained by fitting a Gaussian independently in each angular bin from residual images that can contain imperfect background subtraction, host subtraction artefacts, and contamination (see Figure~B1 in \citetalias{Sola2025}). Instead, the tracks could have been fitted with a spline but this would correlate each angle bin to its neighbours. As a result of our chosen method, the inferred radius can change abruptly between neighbouring bins even when the underlying stream is intrinsically smooth. In contrast, our forward model produces a smooth stream track in projection, so it cannot reproduce arbitrarily large bin to bin variations in $r$ while still matching the overall geometry. When the quoted per bin uncertainties are smaller than the projected stream width, these local discontinuities are not fully absorbed by $\sigma_{\rm data}$, and the likelihood naturally favours $\sigma_{\rm sys}>0$ even for visually good fits. Additional contributions may arise from missing structure in the assumed potential (e.g. neglecting the baryonic disc; see Section~\ref{subsec:Disk_effect}).

\subsection{Population fit}\label{sec:population}

We combine the individual stream fits from the previous section to constrain the parameters of the population distribution of halo flattening. To do so, we adopt the importance sampling approach described in Section~\ref{subsec:population_likelihood}, reweighting the individual posterior samples to evaluate the hierarchical likelihood. Since all non flattening parameters are treated as nuisance parameters, we operate directly on the marginal posteriors of the flattening parameter $q$.

We assume that the underlying population distribution of $q$ follows a truncated Gaussian characterised by mean $\mu_q$ and intrinsic scatter $\sigma_q$. This assumption is motivated by cosmological simulations which predict a cluster distribution of halo shapes around $q\approx0.7$ \citep{Chua2019, Prada2019}. We normalise this truncated Gaussian over its support $q\in[0.5,1.5]$ so that it integrates to unity for every $(\mu_q,\sigma_q)$, ensuring it remains a proper probability density; this prevents the truncation from biasing the population inference toward spherical ($\mu_q=1$) distributions. We adopt uniform priors $\mu_q \in [0.5, 1.5]$ and $\sigma_q \in [0, 0.6]$, restricting the population scatter to the physically relevant range. The hyperparameter inference is performed using dynamic nested sampling with \texttt{dynesty}, starting from 500 live points.

To assess the impact of model mismatch and data quality on the population inference, we divide the stream sample based on the fractional contribution of the systematic term to the total effective uncertainty, 
\[
\frac{\sigma_{\rm sys}}{\sigma_{\rm sys} + \sigma_{\rm data}}.
\]
We define the diagnostic ratio using the standard deviations rather than the variances because our goal is not to measure the exact fractional contribution to the total likelihood, but to obtain an intuitive proxy for how strongly the fit relies on the additional systematic term. The left panel of Figure~\ref{fig:population} shows the distribution of this ratio for the full STRRINGS sample. We adopt a threshold of 70\% to define a \textit{gold} subsample (ratio $<70\%$ with 17 streams) and a \textit{bronze} subsample (ratio $>70\%$ with 15 streams). This threshold is chosen to be roughly in the middle and separating what seems to be a bimodal distribution.

The middle and right panels of Figure~\ref{fig:population} show the resulting posterior distributions of $(\mu_q,\sigma_q)$ for these two subsets. For the \textit{gold} sample, the population inference favours an oblate distribution with $\mu_q = 0.72^{+0.16}_{-0.14}$ and $\sigma_q = 0.34^{+0.18}_{-0.19}$. In this subset, the spherical case $\mu_q=1$ lies above the $\sim1.5\sigma$ level, so spherical and prolate configurations are disfavoured, although not decisively excluded. In contrast, the \textit{bronze} sample yields $\mu_q = 0.97^{+0.19}_{-0.17}$ and $\sigma_q = 0.30^{+0.20}_{-0.16}$, consistent with a spherical distribution but with a markedly larger uncertainty on $\mu_q$ than the \textit{gold} sample. We interpret this not as a genuine preference for spherical halos but as a lack of constraining power. When $\sigma_{\rm sys}$ dominates the effective uncertainty, the projected stream track carries limited geometric information about the underlying potential, allowing a broad range of halo shapes to fit the data. Because the flattening prior is uniform on $[0.5,1.5]$ (symmetric and centred on $q=1$) and the truncated Gaussian population model is properly normalised over this range, the population posterior simply relaxes toward this prior as information is lost, yielding a mean near $\mu_q\approx1$ with a correspondingly inflated uncertainty rather than an artefactual spherical bias.

An alternative interpretation is that the \textit{bronze} sample genuinely traces a nearly spherical halo population. However, the large uncertainties on $\mu_q$ and $\sigma_q$ argue against a confident spherical detection, and this interpretation would not naturally explain why these systems systematically exhibit a larger contribution from the additional systematic term, nor why the \textit{gold} sample prefers an oblate population. The elevated systematic ratios instead suggest that the projected stream tracks in this subsample are less well described by our simplified axisymmetric model, reducing the constraining power of the individual fits and leading to a population inference that is closer to the prior expectation. In this context, it is therefore expected that streams with lower systematic ratios, corresponding to more informative tracks, provide the most reliable constraints on the underlying halo population.

Considering only the \textit{gold} subsample, the inferred oblate population is broadly consistent with expectations from hydrodynamical simulations. For example, IllustrisTNG and Auriga report typical minor-to-major axis ratios around $q \sim 0.7$ in the inner halo \citep{Chua2019,Prada2019}. Our inferred mean flattening of $\sim0.72$ is in good agreement with the $q \sim 0.7$ predicted by these simulations, which lies near the centre of the credible interval of our population model. We emphasise that our modelling neglects the stellar mass component and assumes an axisymmetric NFW potential; the recovered flattening should therefore be interpreted as an effective flattening of the total enclosed potential within the radial range probed by the streams, rather than a direct measurement of the intrinsic dark matter halo shape. In Section~\ref{subsec:Disk_effect}, we quantify how neglecting the disk component can bias the inferred population parameters.

A similar effort to constrain extragalactic halo shapes from streams was recently presented by \citet{Starkman2026}, who modelled the projected morphologies of 13 streams in the \textit{Euclid} Quick Data Release and reported a typical halo shape close to spherical ($q\approx 0.95$). This measurement is not directly comparable to our population inference, given that they directly recover the on-sky potential flattening whereas we forward model the streams in three dimensions. In turn, their approach is considerably faster than our pipeline, so the two methods present a clear trade-off between computational speed and directly constraining the three-dimensional halo. The two analyses are therefore complementary probes of extragalactic halo structure rather than measurements of the same quantity, and a direct comparison of the inferred population values would not be meaningful.

\begin{figure*}
    \centering
    \includegraphics[width=\linewidth]{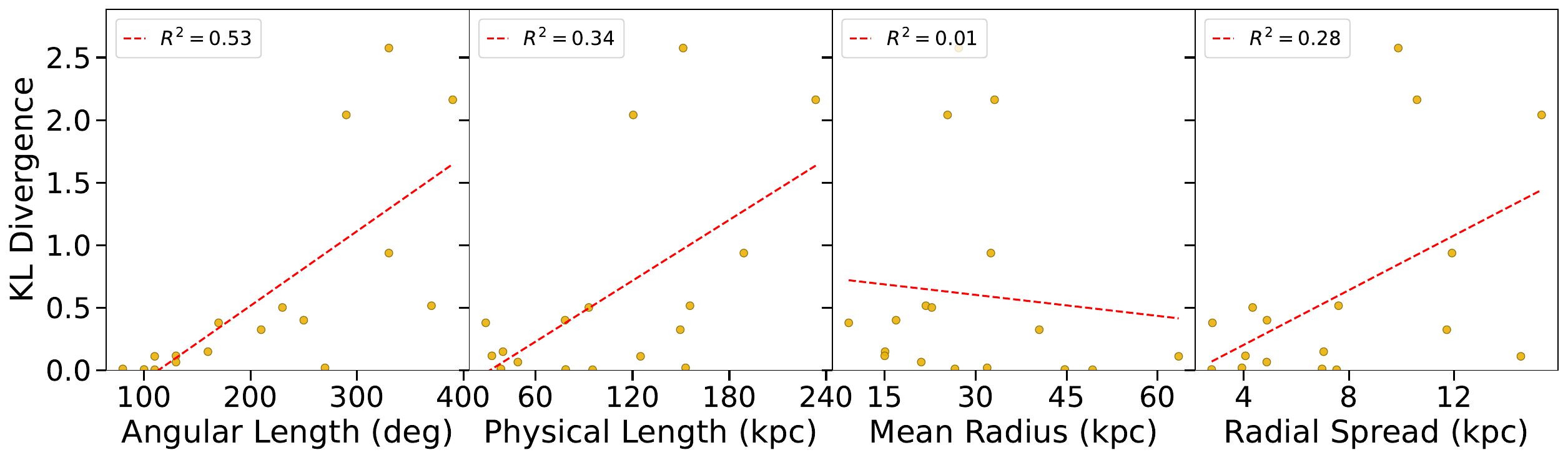}
    \caption{KL divergence between the posterior and the uniform prior on halo flattening, shown for the 17 \textit{gold} streams, as a function of four stream properties (from left to right): the angular length, the physical length, the mean projected radius, and the projected radial spread (the standard deviation of the radial range probed by the projected track). All panels share the same vertical ($D_{\rm KL}$) axis. In each panel the red dashed line is the best fitting linear trend, with its coefficient of determination $R^2$ given in the legend; a larger $R^2$ indicates a tighter correlation, i.e. a more predictive observable. The clearest correlation is with angular length, while the other relations exhibit larger scatter and are likely driven in part by covariance with angular extent.}
    \label{fig:properties}
\end{figure*}

\begin{figure*}
    \centering
    \includegraphics[width=\linewidth]{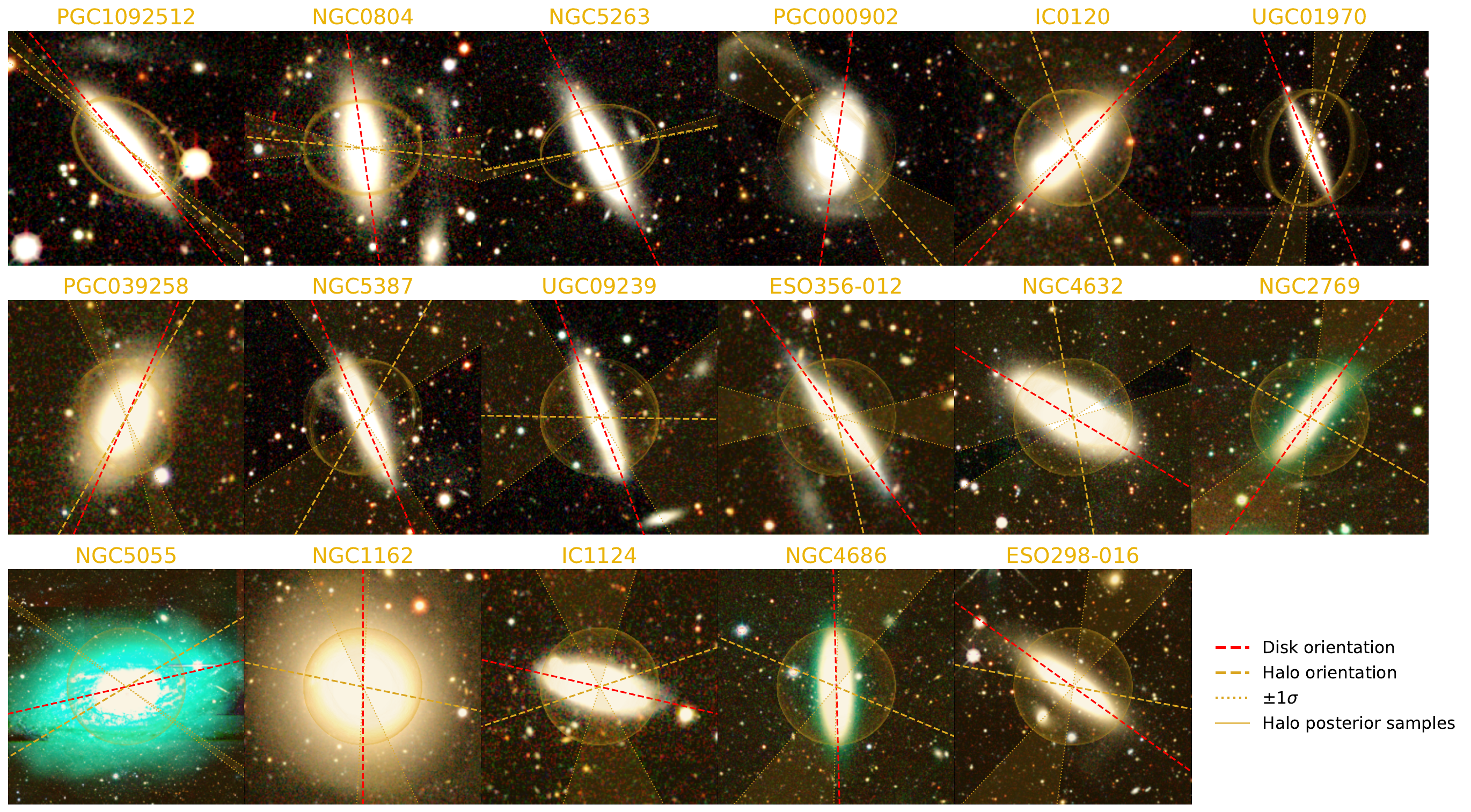}
    \caption{Projected halo orientations for all 17 \textit{gold} galaxies from the STRRINGS catalogue, ordered by constraining power (tightest first). The background images are taken from DESI-LS. The red dashed line shows the disk position angle (PA). The gold ellipses represent $1000$ posterior samples of the projected halo major axis; the dashed gold line is the median orientation, the dotted gold lines mark the 16th and 84th percentiles, and the shaded gold wedge spans this $1\sigma$ range. The size of the ellipses was arbitrarily chosen to be slightly wider than the galaxy for plotting reasons and is not to scale. The legend in the final panel summarises these elements.}
    \label{fig:alignment}
\end{figure*}

\begin{figure}
    \centering
    \includegraphics[width=\linewidth]{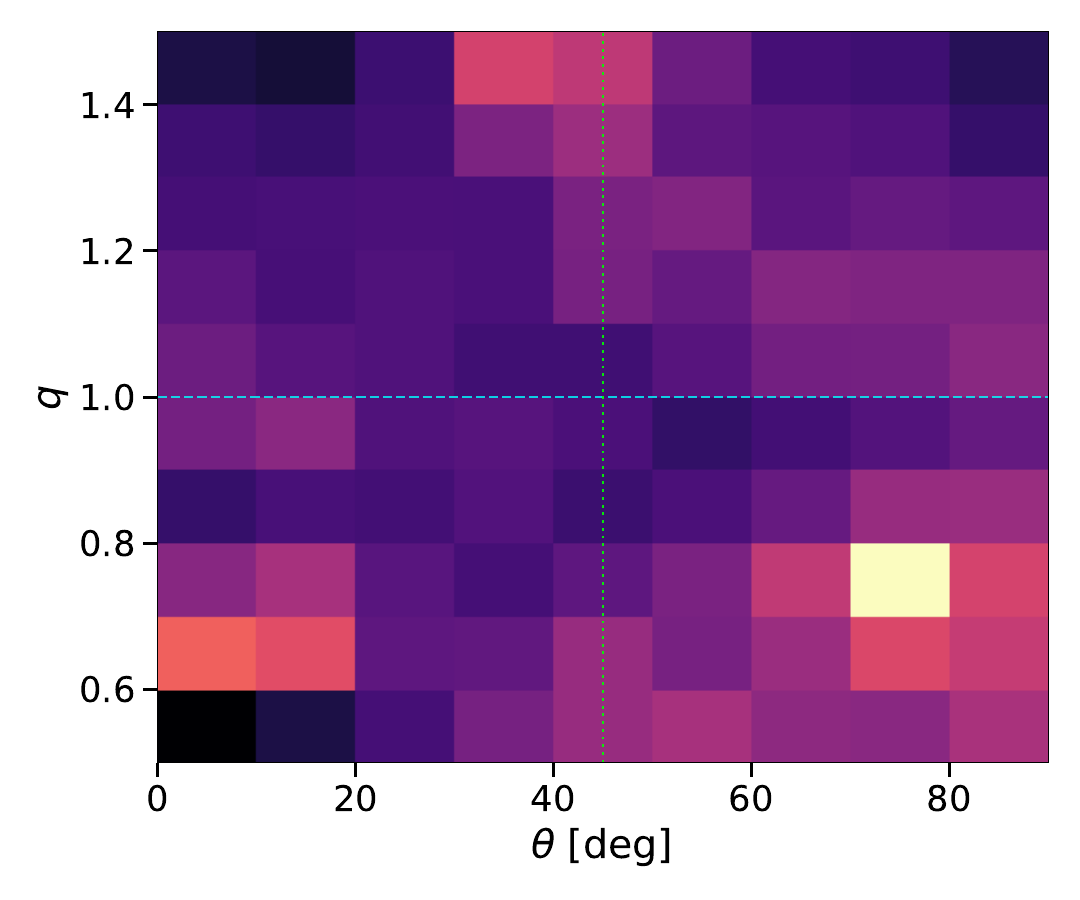}
    \caption{Joint probability density distribution of the projected misalignment angle $\theta$ (between the disk PA and the halo \emph{major} axis) and the halo flattening $q$, pooled over all 17 \textit{gold} streams. Each stream contributes an equal number of posterior samples, so every galaxy carries equal weight; the pooled $(\theta,q)$ samples are binned in $10^\circ\times0.1$ cells and the colour shows the (power-stretched) density. The cyan dashed line marks the spherical case $q=1$ (oblate below, prolate above) and the green dotted line $\theta=45^\circ$. Three over-densities (``islands'') stand out: oblate haloes that are nearly \emph{aligned} ($\theta\approx0$--$20^\circ$, $q\approx0.6$--$0.7$) or nearly \emph{perpendicular} ($\theta\approx70$--$80^\circ$, $q\approx0.7$--$0.8$), and prolate haloes at \emph{intermediate} angles ($\theta\approx30$--$50^\circ$, $q\gtrsim1.4$). Figure~\ref{fig:theta_dist} collapses this onto the $\theta$ axis.}
    \label{fig:theta_q_2d}
\end{figure}

\begin{figure}
    \centering
    \includegraphics[width=\linewidth]{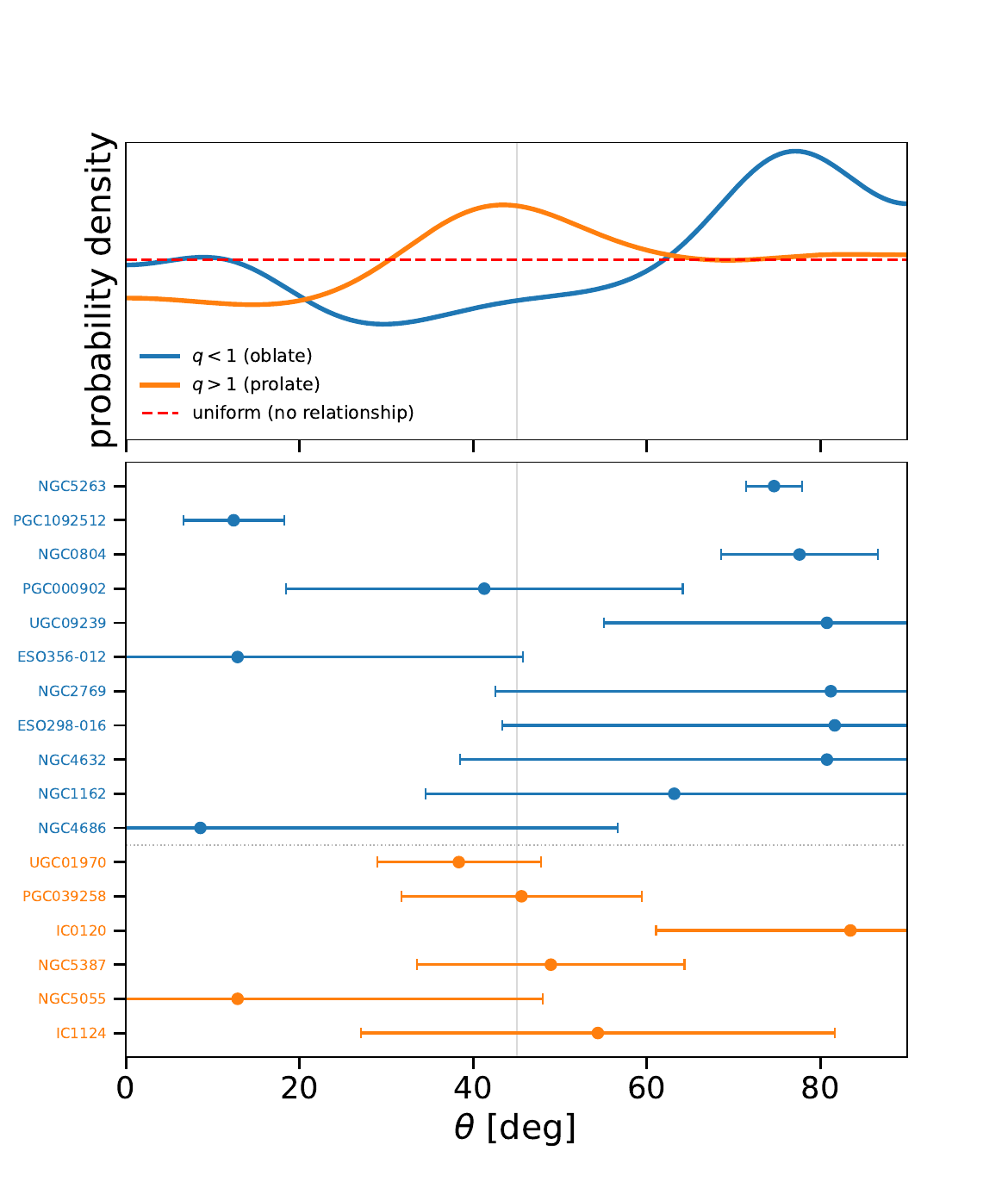}
    \caption{Projected misalignment angle $\theta$ between the disk position angle (PA) and the halo \emph{major} axis, split by halo shape ($\theta=0^\circ$ for alignment, $\theta=90^\circ$ for perpendicularity). \emph{Top:} the population-level distribution of $\theta$, built by pooling the posteriors of all 17 \textit{gold} streams --- drawing an equal number of samples from each so that every galaxy contributes with equal weight --- and smoothing with a kernel density estimate (KDE), shown separately for oblate ($q<1$, blue) and prolate ($q>1$, orange) haloes. The red dashed line is the flat $\theta\sim\mathcal{U}[0^\circ,90^\circ]$ expectation for no disk--halo correlation. \emph{Bottom:} the individual fits for all 17 streams, each shown as the maximum-a-posteriori $\theta$ (point) with a symmetric $65\%$ credible interval (error bar) measured from a KDE of that stream's $\theta$ posterior. Streams are ordered by constraining power --- the KL divergence of their $\theta$ posterior from a uniform distribution --- with the oblate streams listed first (most to least constrained) and the prolate streams below the dotted separator. The faint vertical line marks $\theta=45^\circ$.}
    \label{fig:theta_dist}
\end{figure}

\section{Discussion}\label{sec:discussion}

\subsection{Constraints based on stream properties}\label{subsec:properties}

We quantify how informative each \textsc{STRRINGS} stream is for constraining the halo flattening, $q$, by relating the constraining power of the individual posterior to observable stream properties. As an information metric, we use the Kullback--Leibler (KL) divergence between the marginal posterior on the flattening, $p(q\mid d)$, and the corresponding uniform prior, $p_0(q)$:
\begin{equation}
D_{\rm KL}\!\left[p(q\mid d)\,\|\,p_0(q)\right]
= \int_{q} p(q\mid d)\,\log\!\left[\frac{p(q\mid d)}{p_0(q)}\right]\,{\rm d}q.
\label{eq:kl_def}
\end{equation}
Here, $D_{\rm KL}$ is interpreted as the information gain provided by the data relative to the prior: $D_{\rm KL}=0$ corresponds to a posterior identical to the prior (no constraining power), while larger values indicate increasingly informative posteriors.

In practice, we estimate $D_{\rm KL}$ from posterior samples using a histogram approximation. We adopt a common set of bin edges for all streams to ensure comparability, using 30 uniform bins over $q\in[0.5,1.5]$ (the range of the uniform prior); we verified that the qualitative trends discussed below are stable to reasonable changes in binning. Let $\hat p_j$ be the posterior probability mass in bin $j$ and $p_{0_j}$ the corresponding prior mass (constant for a uniform prior). We then compute
\begin{equation}
D_{\rm KL} \approx \sum_{j} \hat p_j \,\log\!\left(\frac{\hat p_j}{p_{0,j}}\right),
\label{eq:kl_hist}
\end{equation}
with the convention that bins with $\hat p_j=0$ contribute zero.

Figure~\ref{fig:properties} shows the resulting KL divergences for the 17 \textit{gold} streams as a function of four stream properties: angular length, physical length, mean projected radius, and radial spread (from left to right). We restrict this comparison to the \textit{gold} sample because the \textit{bronze} fits have a big mismatch with our model which might bias our analysis. We define the observables as follows: the angular length is the total polar angle extent $\Delta\phi$ spanned by the track; the physical length is the projected on-sky length of the track in kpc; the mean projected radius is the track averaged projected distance $\langle r\rangle$ from the host centre; and the radial spread is the standard deviation of the projected radius $r$ along the track. These are the same quantities explored in \citetalias{Chemaly2026}, where constraining power could be assessed directly against ground truth using mock catalogues (see Appendix A of that paper). Here, we find qualitatively similar behaviour. To quantify the constraining power associated with each property, we report the coefficient of determination, $R^2$, of a linear fit between $D_{\rm KL}$ and that property (shown in the legend of each panel of Figure~\ref{fig:properties}); a larger $R^2$ reflects a tighter correlation and hence a more predictive observable. The strongest correlation is with the angular length ($R^2\simeq0.53$): longer tracks yield larger $D_{\rm KL}$ values, i.e. more informative posteriors. This is expected because the likelihood is evaluated over angular bins, so a larger angular span typically provides more independent constraints on the projected geometry \citep{Walder2024}.
Motivated by the demonstrated sensitivity of stream curvature to the shape of the host potential \citep{Nibauer2023,Nibauer2025b,Wu2026}, we also explored whether the curvature of the projected track is a useful predictor of constraining power. Because the raw \textsc{STRRINGS} tracks are not smooth enough to differentiate reliably, we estimated the curvature following the smoothing-spline procedure of \citet{Wu2026} and \citet{Starkman2026}, fitting a smooth curve to each track and normalising by the track length so as to isolate the intrinsic bendiness from the sheer angular extent. We find no clear trend between this curvature and the constraining power. A possible explanation might be the lack of dynamical range: the \textsc{STRRINGS} tracks span only a narrow range of intrinsic curvature ($\langle|\kappa|\rangle \approx 0.07 \pm 0.04\,{\rm kpc}^{-1}$), which is unsurprising given that the catalogue was assembled, in part, by preferentially selecting long and visibly curved streams. With such limited diversity in curvature, a genuine dependence is hard to isolate; establishing one would require a larger and more varied sample spanning a broader range of stream shapes.

We also observe positive correlations between $D_{\rm KL}$ and both the physical length ($R^2\simeq0.36$) and the radial spread ($R^2\simeq0.27$). These relations still exhibit more scatter than the angular length and are likely driven in part by covariance with angular extent (longer streams tend to be long in multiple metrics). Finally, the mean projected radius is essentially uninformative ($R^2\simeq0.01$), with at most a weak apparent anti-correlation. We do not expect the sensitivity to halo flattening to depend directly on radius in a simple monotonic way for projected tracks, and we therefore interpret any residual trend primarily as a geometric selection effect: at smaller projected radii, a given physical length subtends a larger angular extent, which again increases the amount of information available in the track.

Overall, the angular length emerges as the most predictive single observable for the constraining power of a photometry only stream track in our framework given our dataset, with the largest $R^2$ of the four properties considered, while the other properties provide at most secondary predictive value once their covariance with angular extent is accounted for.

\subsection{Projected alignment between the disk and halo}

We briefly investigate the projected alignment between the stellar disk and the inferred halo orientation. The position angle (PA) of the galaxy disk is taken directly from SGA-2020 \citep{Moustakas2023}, while the halo orientation is obtained from the posterior distribution of the stream model. Because the inclinations of most host galaxies are not reliably constrained, we restrict this analysis to projected quantities only; this therefore does not constitute a measurement of the true three dimensional alignment between disk and halo.

Figure~\ref{fig:alignment} shows, for all 17 \textit{gold} galaxies, the projected halo orientation inferred from the stream model overlaid on the DESI-LS image, together with the disk PA. For each galaxy the gold ellipses trace $1000$ posterior samples of the projected halo major axis, the dashed gold line marks its median orientation and the dotted gold lines the 16th--84th percentiles, while the red dashed line is the disk PA. The projected halo major axis is taken to be the model symmetry axis itself when the halo is prolate ($q>1$) and its $90^\circ$ rotation when it is oblate ($q<1$). The width of the orientation wedge is directly tied to how well the flattening is constrained. When the posterior is consistent with a spherical halo ($q\approx1$) or leaves the flattening largely unconstrained, the halo has essentially no preferred axis: the projected major axis becomes ill defined and the posterior samples spread over a wide range of position angles, producing a broad $1\sigma$ wedge and a nearly isotropic (spherical) distribution of sampled orientations. Conversely, when the stream tightly constrains the flattening to be clearly oblate or prolate, the symmetry axis is well defined and the inferred orientation shows a clear, narrow preference.

Inspecting the individual fits, the projected halo major axis spans a wide range of orientations relative to the disk PA. To search for structure at the population level, Figure~\ref{fig:theta_q_2d} shows the joint distribution of the projected misalignment angle $\theta$ (between the disk PA and the halo major axis, with $\theta=0^\circ$ for alignment and $\theta=90^\circ$ for perpendicularity) and the halo flattening $q$. We build this by pooling the posteriors of all 17 \textit{gold} streams, drawing an equal number of samples from each so that every galaxy contributes with equal weight, and binning the pooled $(\theta,q)$ samples in $10^\circ\times0.1$ cells. Rather than a single diffuse blob, the distribution breaks into three distinct over-densities, which we refer to as ``islands'': oblate haloes ($q<1$) that are nearly aligned ($\theta\approx0$--$20^\circ$), oblate haloes that are nearly perpendicular ($\theta\approx70$--$80^\circ$), and prolate haloes ($q>1$) at intermediate angles ($\theta\approx30$--$50^\circ$).

Collapsing this joint distribution onto the $\theta$ axis sharpens the islands into clear modes (Figure~\ref{fig:theta_dist}). We smooth the pooled $\theta$ samples with a kernel density estimate (top panel) and, motivated by the two distinct shape regimes, show the oblate ($q<1$) and prolate ($q>1$) haloes separately. The bottom panel shows the corresponding per-stream measurements: for each stream we report the maximum-a-posteriori $\theta$ with a symmetric $65\%$ credible interval, both obtained from a KDE of that stream's $\theta$ posterior, and we order the streams by how tightly $\theta$ is constrained, quantified by the KL divergence of the per-stream $\theta$ posterior from a uniform distribution (computed as in Equation~\ref{eq:kl_def}, here for $\theta$ against $\mathcal{U}[0^\circ,90^\circ]$).

The two populations behave strikingly differently. The oblate haloes pile up towards the extremes of the range, with excess probability near alignment ($\theta\to0^\circ$) and perpendicularity ($\theta\to90^\circ$) and a clear deficit at intermediate angles. A preference for alignment has a natural dynamical explanation: a baryonic disk is most stable when it settles into the equatorial plane of an oblate halo (perpendicular to the halo symmetry axis), which projects to alignment between the disk PA and the projected halo major axis. The prolate haloes, in contrast, favour neither extreme but instead peak at intermediate angles ($\theta\approx30$--$50^\circ$). Such an oblique preference is not naturally accounted for by a simple stability argument, and if it is genuine it may point to a more fundamental aspect of how disks settle within prolate or triaxial potentials. We regard this only as a tentative indication.

These trends should be interpreted with caution. As the bottom panel makes clear, the population-level signal is dominated by a handful of tightly constrained streams: only the few highest-KL systems have narrow error bars, while the majority are essentially unconstrained and span a broad range of $\theta$. The measurement is also purely projected --- the host inclinations are not constrained, so an apparent (mis)alignment on the sky need not reflect the true three dimensional geometry --- and each stream probes the potential only over a limited radial range. Cosmological simulations moreover predict that disk--halo alignment is primarily a three dimensional, radius dependent effect, strongest between the stellar disk and the inner halo with increasing scatter outwards \citep[e.g.][]{Chua2019,Prada2019}, which projected position angles alone cannot test. A larger sample of well-constrained streams will be required to confirm whether these shape-dependent trends are genuine statistical signals at the population level.

A full assessment therefore requires modelling the disk and halo simultaneously in three dimensions, which will be the subject of future work. Nevertheless, these fits demonstrate that the stream tracks contain measurable information about the projected halo orientation.

\begin{figure}
    \centering
    \includegraphics[width=\linewidth]{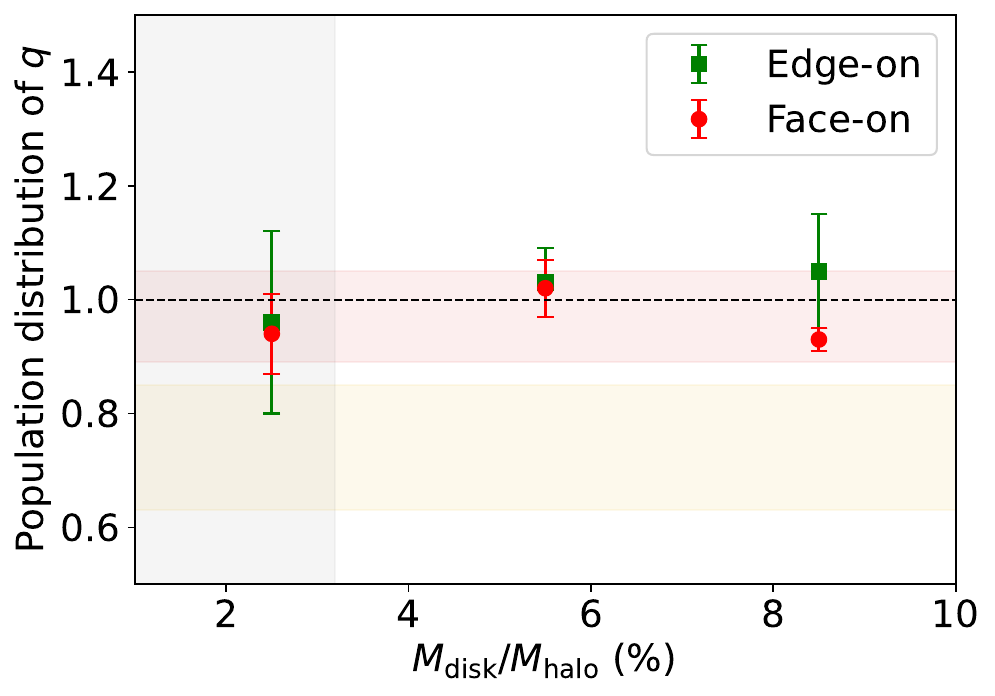}
    \caption{Recovered population parameters $(\mu_q, \sigma_q)$ as a function of disk-to-halo mass ratio for mock populations including a Miyamoto–Nagai disk but analysed with a disk free halo model. Each point represents a population of 32 streams. Edge-on (green) and face-on (red) disk orientations are shown separately. The shaded regions indicate the $1\sigma$ constraints from the \textit{gold} (in yellow) and \textit{bronze} (in red) samples in the real \textsc{STRRINGS} data. The grey region marks the $1\sigma$ disk-to-halo mass ratio inferred for the \textsc{STRRINGS} hosts. A significant oblate bias appears only for the most massive, face-on configuration (rightmost red point); at the disk-to-halo mass ratios representative of the \textsc{STRRINGS} hosts no significant bias is detected.}
    \label{fig:pop_disk}
\end{figure}
 
\subsection{Effect of the missing disk}\label{subsec:Disk_effect}

Our modelling framework does not explicitly include a baryonic disk component. To quantify the potential impact of this omission on the inferred population distribution of halo flattening, we construct controlled mock populations that include a stellar disk but are analysed with our missing disk pipeline.

We generate six mock populations, each containing 32 streams. For each population, the host potential consists of a spherical NFW halo combined with a Miyamoto–Nagai disk \citep{Miyamoto1975}. The halo parameters are drawn from the same priors adopted in the individual fits including a $\sigma_{sys}$. The disk orientation is fixed to be either face-on or edge-on, and for each orientation we consider three disk-to-halo mass ratio ranges sampled uniformly from 1–4\%, 4–7\%, and 7–10\%. In total, this yields over 150 mock streams. The disk scale length and scale height are fixed to 3.5\,kpc and 0.5\,kpc, respectively, comparable to slightly larger than Milky Way values. For reference, using stellar masses from DECaLS photometry and the stellar-to-halo mass relation from \citet{Moster2013} at $z\sim0$, we estimate that the average disk-to-halo mass ratio in the \textsc{STRRINGS} sample is $\sim 2\%$.

Each mock stream is projected, binned in fixed angular intervals, and Gaussian radial noise of 2\% is added to match the typical level in \textsc{STRRINGS} (same procedure as \citetalias{Chemaly2026}). We then perform inference assuming an axisymmetric halo without a disk, exactly as in the main analysis, and combine the individual posteriors via importance sampling to recover the population parameters $(\mu_q, \sigma_q)$ under a Gaussian model. If the disk contribution does not bias our inference, the recovered mean flattening should remain consistent with $q=1$, since the underlying halo in the mocks is spherical.

Figure~\ref{fig:pop_disk} shows the recovered population parameters as a function of disk-to-halo mass ratio for both edge-on (green) and face-on (red) configurations. Each point represents the result of one 32 stream population fit. For the edge-on configurations, and for the face-on configurations at lower disk masses, the inferred mean flattening remains consistent with the spherical case ($\mu_q=1$). The one clear exception is the most massive face-on configuration, with a disk-to-halo mass ratio of $7$--$10\%$ (the rightmost red point), for which $\mu_q=1$ is no longer recovered within $1\sigma$: a genuine oblate bias appears once the disk is massive enough to imprint on the projected tracks.

This face-on/edge-on asymmetry follows naturally from how our mock streams are selected. By construction, the mock tracks are required to span a similar projected angular length to the observed \textsc{STRRINGS} streams. For an edge-on disk, a stream lying in the disk plane is strongly foreshortened in projection and is therefore unlikely to satisfy this angular length selection, so the surviving streams rarely probe the plane where the disk's influence is largest. For a face-on disk, by contrast, long and well-curved projected tracks can lie within the disk plane, precisely where the disk contributes most to the potential; these streams are then influenced enough to bias the recovered population flattening. The bias therefore emerges only when the disk is both massive and seen face-on. Reassuringly, the disk-to-halo mass ratios typical of the \textsc{STRRINGS} hosts ($\sim2\%$) lie well below this regime, so our inferred flattening distribution is unlikely to be driven by the omission of a modest baryonic disk component. We note that these mock fits retain the same additional systematic term $\sigma_{\rm sys}$ as the main analysis: the disk-free model cannot perfectly reproduce the disk-bearing mock tracks, so the model and data do not exactly align, and we keep $\sigma_{\rm sys}$ in order to mirror our \textsc{STRRINGS} pipeline as closely as possible.

We emphasise, however, that these tests rely on a simplified set of assumptions and are not intended to provide a fully realistic representation of the \textsc{STRRINGS} sample. In particular, the adopted mock populations do not reproduce the full diversity of host properties, viewing geometries, or stream morphologies present in the data. To partially address this, we performed additional tests in which the configuration space of the projected mock tracks was chosen to better resemble that of the observed \textsc{STRRINGS} streams, and again found no clear evidence that the omission of a disk introduces a systematic bias in the recovered population flattening. Nevertheless, other disk-related effects not captured by these simplified experiments may still be important. A full treatment therefore requires incorporating an explicit disk component directly into the stream fits, which we leave to future work. For the purposes of the present analysis, these tests suggest that the impact of neglecting a baryonic disk is likely to be subdominant.

\section{Conclusions}\label{sec:conclusions}

We have presented an empirical population level inference of dark matter halo flattening from stellar streams beyond the Local Group. Building on the hierarchical framework developed in \citetalias{Chemaly2026}, we forward modelled 32 streams from the \textsc{STRRINGS} catalogue \citetalias{Sola2025} using an axisymmetric NFW potential and fit each system using only the projected stream track. We then combined the resulting individual posteriors through importance sampling to infer the underlying population distribution of halo flattening.

Individual stream constraints span a wide range of information content. Several systems remain weakly constraining or exhibit projection driven multi-modality, while a subset yields tight posteriors on flattening. To diagnose cases where the fit is dominated by model mismatch or track systematics, we introduced an additional variance term, $\sigma_{\rm sys}$, and used the fractional contribution of that parameter to separate a \textit{gold} and \textit{bronze} subsample. The \textit{gold} subset contains the streams for which the projected track retains significant information about the potential.

Our main results are summarised as follows:
\begin{enumerate}
    \item We obtain posterior constraints on halo flattening for 32 \textsc{STRRINGS} systems using only stream tracks from photometric data. The forward model reproduces the dominant stream geometries in projection, while the inferred $\sigma_{\rm sys}$ indicates a typical additional scatter of a few kpc, consistent with modest model mismatch and track extraction systematics.
    \item Individual posteriors often remain broad and can be multi-modal between oblate and prolate solutions, reflecting known projection degeneracies for extragalactic streams. Nevertheless, the ensemble of posteriors contains sufficient information to constrain population level parameters.
    \item Combining the individual posteriors under a truncated Gaussian population model yields a mildly oblate distribution of halo flattening. For the \textit{gold} subsample (17 streams), we infer $\mu_q = 0.72^{+0.16}_{-0.14}$ and intrinsic scatter $\sigma_q = 0.34^{+0.18}_{-0.19}$, with spherical configurations not decisively excluded but both spherical and prolate populations disfavoured beyond the $1.5\sigma$ level.
    \item The \textit{bronze} subsample (15 streams) yields a nearly spherical population inference with substantially larger uncertainty ($\mu_q = 0.97^{+0.19}_{-0.17}$, $\sigma_q = 0.30^{+0.20}_{-0.16}$), consistent with reduced constraining power when $\sigma_{\rm sys}$ dominates the effective uncertainty rather than indicating a distinct physical population.
\end{enumerate}

Interpreting these constraints requires care. Our model assumes an axisymmetric NFW halo and neglects the baryonic disk, so the recovered flattening should be viewed as an effective flattening of the total potential within the radial range probed by the streams, rather than a direct measurement of the intrinsic dark matter halo axis ratio. Within this interpretation, the inferred oblate tendency is broadly consistent with cosmological hydrodynamical simulations that predict inner haloes to be rounder and more oblate when baryons are included, albeit our inferred mean is closer to spherical.

This work establishes a practical pathway for statistical halo shape measurements from external streams using photometry alone. The main limitations of the present analysis motivate clear extensions:
\begin{enumerate}
    \item We fit only the stream track in projection. Incorporating width and surface brightness information will require a more realistic forward model and likelihood, but offers a direct route to stronger constraints.
    \item One projected progenitor coordinate is fixed and progenitor identification remains uncertain in many systems; allowing the progenitor to move freely along the track should reduce modelling systematics in the inner stream.
    \item The host potential is intentionally simplified (axisymmetric NFW). Allowing baryonic components and more general halo morphologies (e.g.\ triaxiality or radial shape variation) will introduce additional physical degeneracies but is necessary for interpreting $q$ as a dark matter specific shape.
    \item Track extraction systematics and residual contamination can degrade constraining power, as reflected by high $\sigma_{\rm sys}$ cases. More automated quality metrics and improved track measurements will be essential for future large samples.
\end{enumerate}

With forthcoming deep imaging from \textit{Euclid} and Rubin/LSST expected to deliver vastly larger samples of extragalactic stellar streams, the hierarchical approach adopted here provides a scalable foundation for precision, population level tests of halo morphology and galaxy formation beyond the Milky Way.

\section*{Acknowledgements}

DC acknowledges funding from the Harding Distinguished Postgraduate Scholars Program. ES is grateful to the Leverhulme Trust for funding under the grant number RPG-2021-205. SK acknowledges support from the Science $\&$ Technology Facilities Council (STFC) grant ST/Y001001/1.

\section*{Data Availability}

The full fitting pipeline, together with the scripts to reproduce all figures and results in this paper, is available upon request.
 


\bibliographystyle{mnras}
\bibliography{references} 




\appendix

\section{Parameter Priors}\label{app:A}

Table~\ref{table:prior_orbit} summarises the prior distributions adopted for the thirteen parameters required to forward model the projected stream track, together with the additional systematic term. These priors are used both in Subsection~\ref{sec:individual} for fitting the individual \textsc{STRRINGS} streams and in Subsection~\ref{subsec:Disk_effect} for fitting the mock streams generated to assess the impact of neglecting the disk component.

The priors are chosen to be broad and minimally informative, while remaining physically plausible based on simulations and observational constraints. The imposed inequality constraints implement the symmetry breaking choices described in \citet{Chemaly2026}.

\begin{table}
\centering
\caption{Priors for the fourteen inferred parameters. $\mathcal{U}(a,b)$ denotes a uniform prior on $[a,b]$; $\mathcal{N}(\mu,\sigma)$ denotes a Gaussian prior with mean $\mu$ and standard deviation $\sigma$.
}
\label{table:prior_orbit}
\resizebox{\linewidth}{!}{%
\begin{tabular}{llll}
\hline
Parameter & Symbol & Prior & Units \\
\hline
Halo mass & $\log_{10}(M/M_\odot)$ & $\mathcal{U}(11,14)$ & -- \\
Halo scale radius & $R_s$ & $\mathcal{U}(10,25)$ & kpc \\
Halo orientation and shape & $\hat{x}, \hat{y}, \hat{z}$ & $\mathcal{N}(0,1)$ with $\hat{z}\ge 0$ & -- \\
Progenitor mass & $\log_{10}(m/M_\odot)$ & $\mathcal{U}(7,9)$ & -- \\
Progenitor scale radius & $r_s$ & $\mathcal{U}(1,5)$ & kpc \\
Final positions & $x_0, z_0$ & $\mathcal{N}(0,150)$ with $x_0>0$, $z_0>0$ & kpc \\
Final velocities & $v_{x_0}, v_{y_0}, v_{z_0}$ & $\mathcal{N}(0,250)$ with $v_{y_0}>0$ & km\,s$^{-1}$ \\
Integration time & $\text{time}$ & $\mathcal{U}(1,4)$ & Gyr \\
Systematic term & $\sigma_{\rm sys}$ & $\mathcal{U}(0,25)$ & kpc \\
\hline
\end{tabular}
}
\end{table}

\section{STRRINGS fits}\label{app:B}

In this appendix, we present the best fitting models and posterior distributions of the halo flattening parameter for the remaining 24 streams out of the 32 successfully fitted systems in the \textsc{STRRINGS} catalogue (see Figures~\ref{fig:AppB1} and~\ref{fig:AppB2}). The other eight representative examples are shown in Figure~\ref{fig:example_individual} in the main text.

The colour of each stream label denotes its sub-sample classification: goldenrod for the \emph{gold} sample and sienna for the \emph{bronze} sample, defined according to the relative contribution of $\sigma_{\rm sys}$ to the total noise budget.

Overall, the best fitting models closely follow the observed stream tracks, supporting the robustness and reliability of our automated fitting procedure.

\begin{figure*}
    \centering

    \begin{minipage}{0.48\textwidth}
        \centering
        \includegraphics[width=\linewidth]{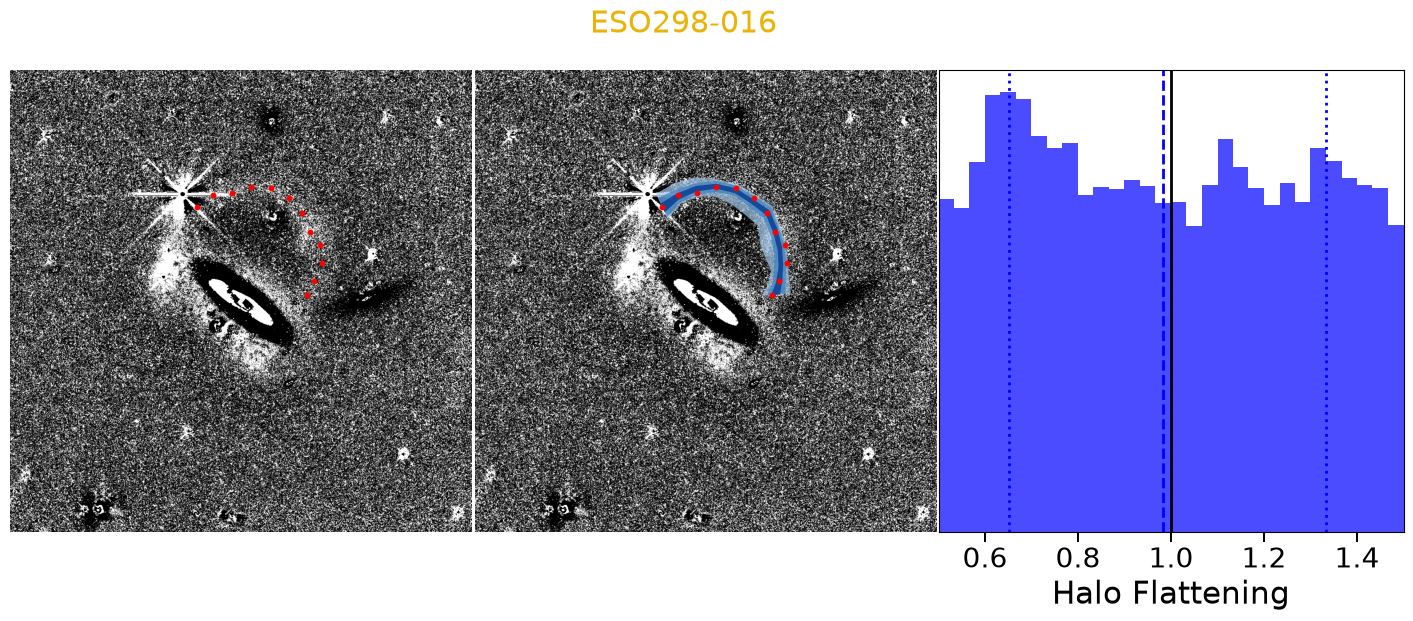}
    \end{minipage}
    \begin{minipage}{0.48\textwidth}
        \centering
        \includegraphics[width=\linewidth]{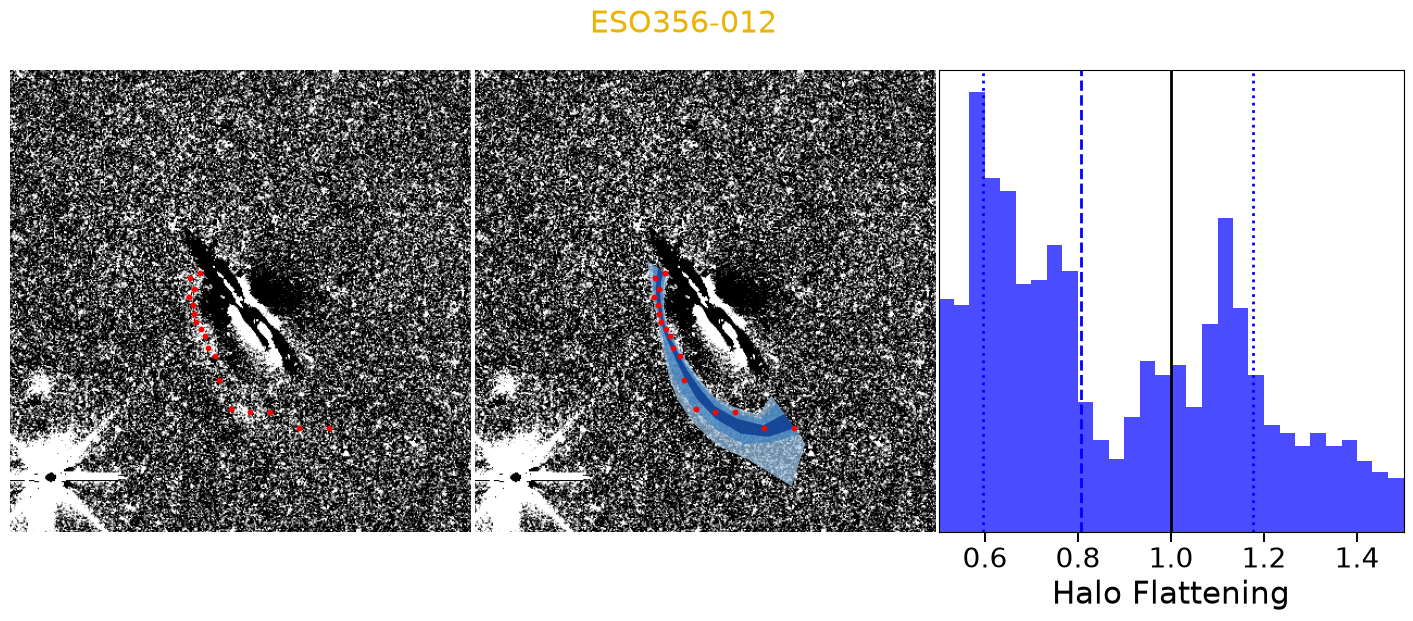}
    \end{minipage}

    \begin{minipage}{0.48\textwidth}
        \centering
        \includegraphics[width=\linewidth]{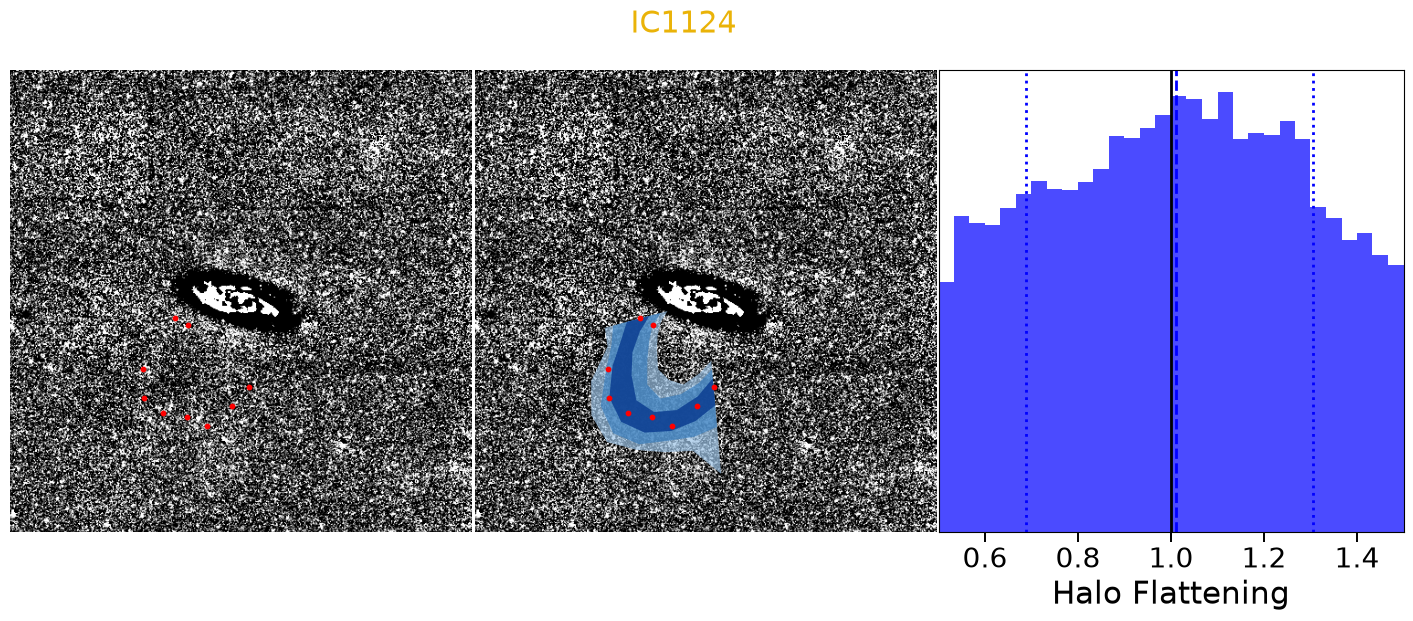}
    \end{minipage}
    \begin{minipage}{0.48\textwidth}
        \centering
        \includegraphics[width=\linewidth]{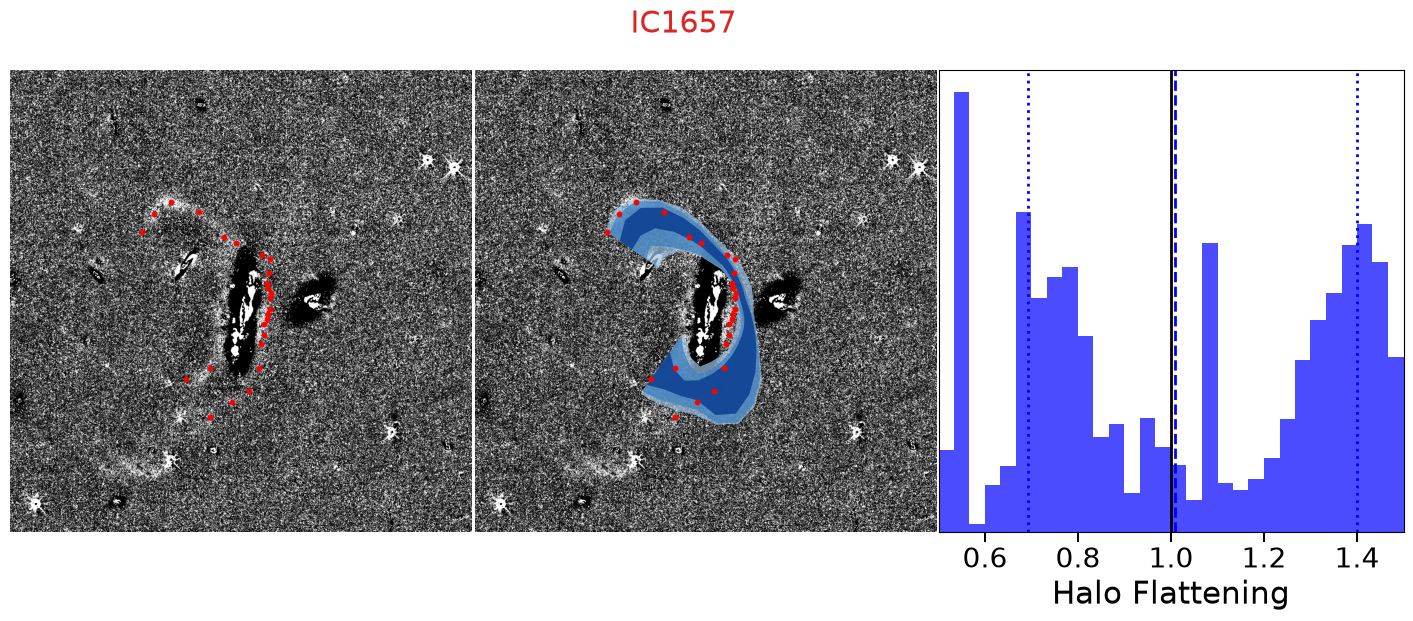}
    \end{minipage}

    \begin{minipage}{0.48\textwidth}
        \centering
        \includegraphics[width=\linewidth]{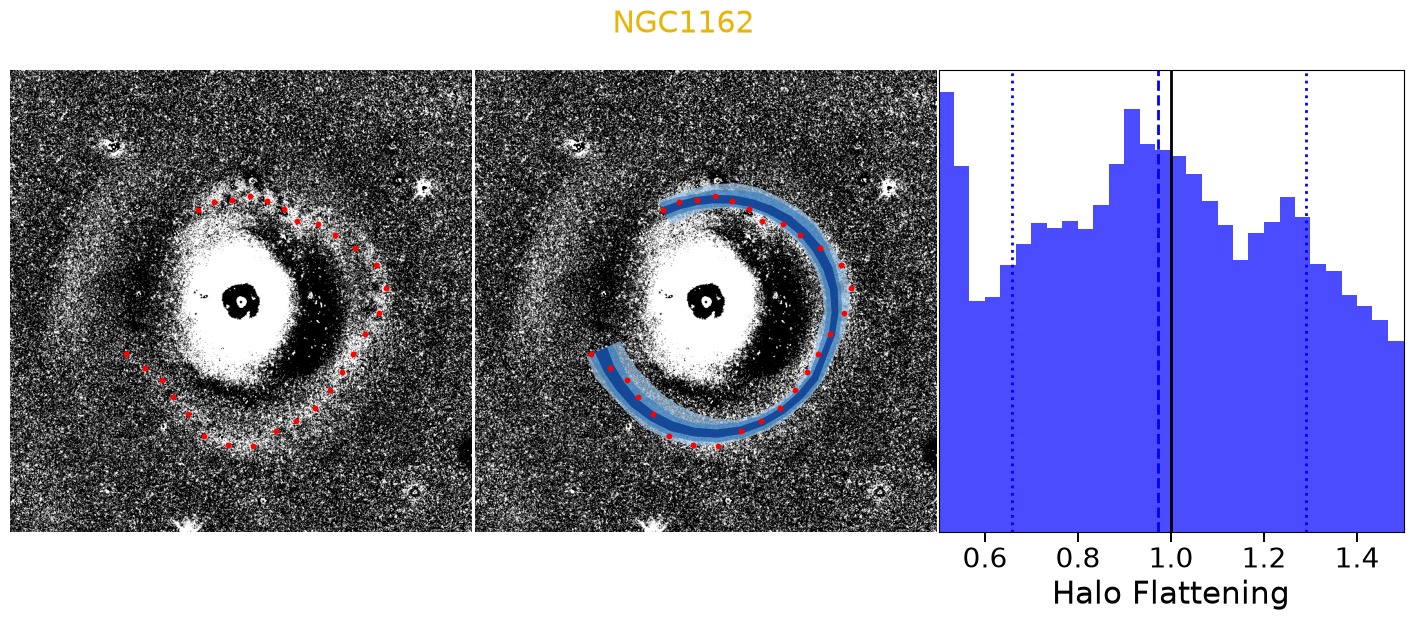}
    \end{minipage}
    \begin{minipage}{0.48\textwidth}
        \centering
        \includegraphics[width=\linewidth]{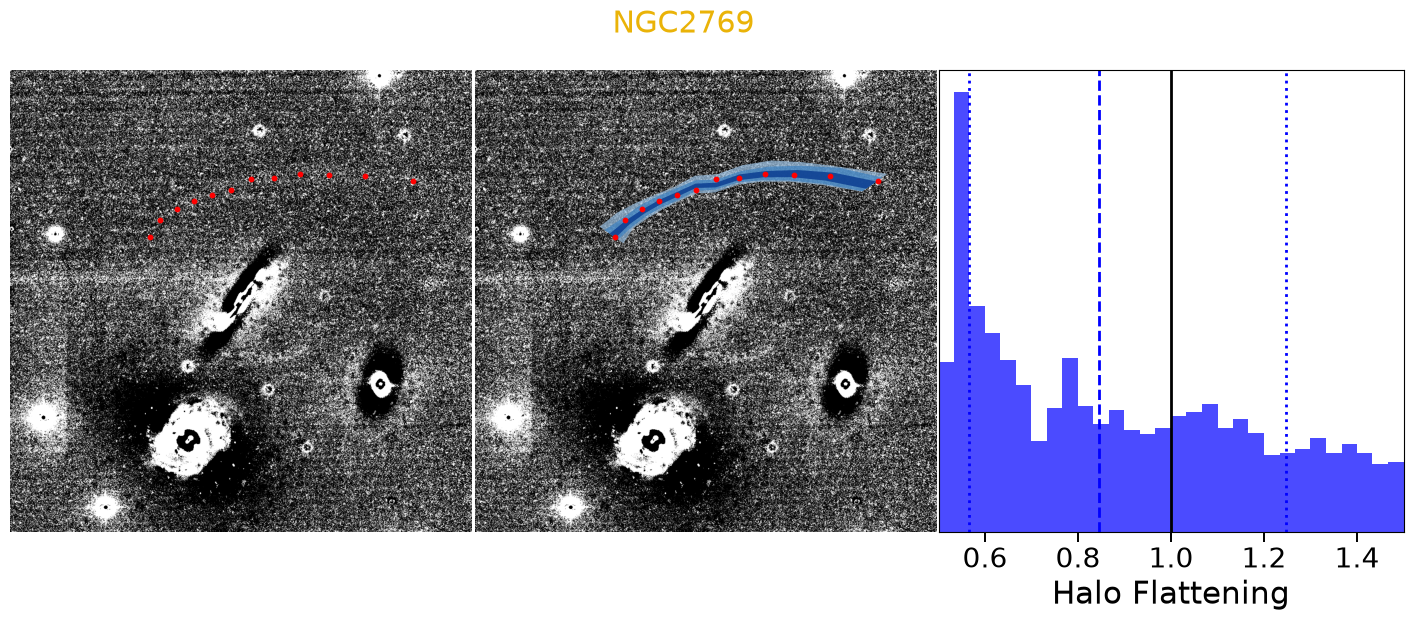}
    \end{minipage}

    \begin{minipage}{0.48\textwidth}
        \centering
        \includegraphics[width=\linewidth]{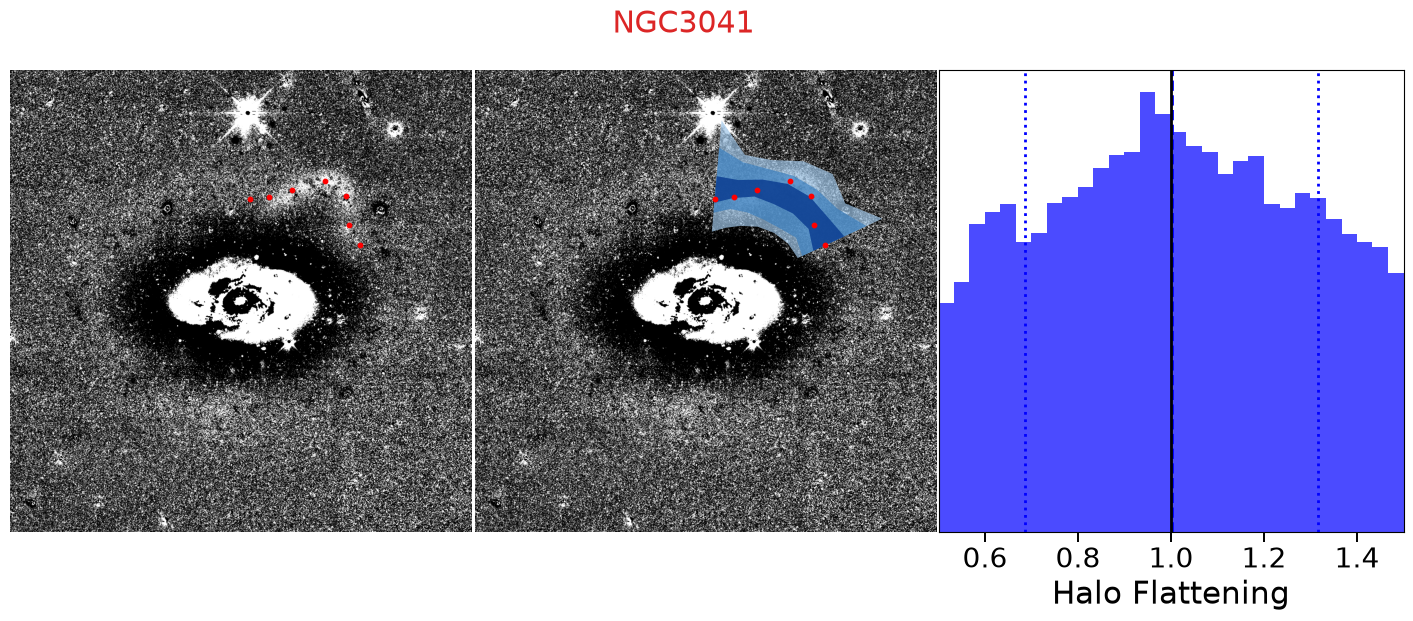}
    \end{minipage}
    \begin{minipage}{0.48\textwidth}
        \centering
        \includegraphics[width=\linewidth]{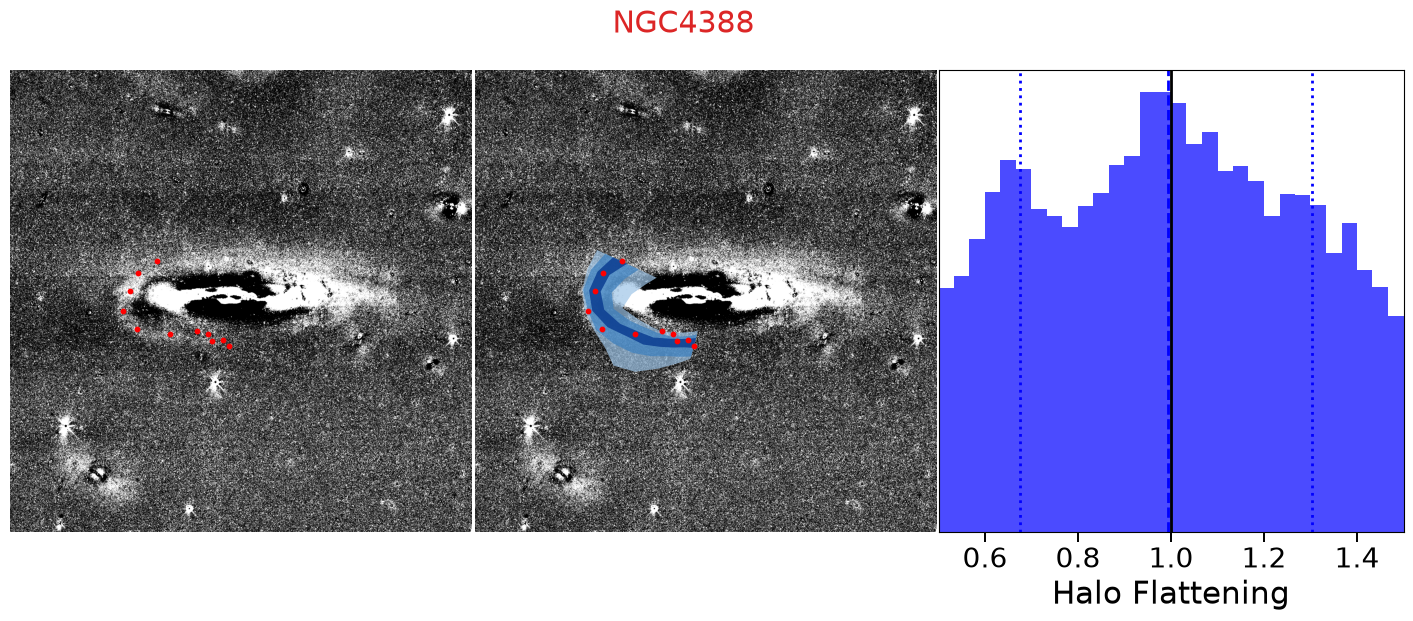}
    \end{minipage}

    \begin{minipage}{0.48\textwidth}
        \centering
        \includegraphics[width=\linewidth]{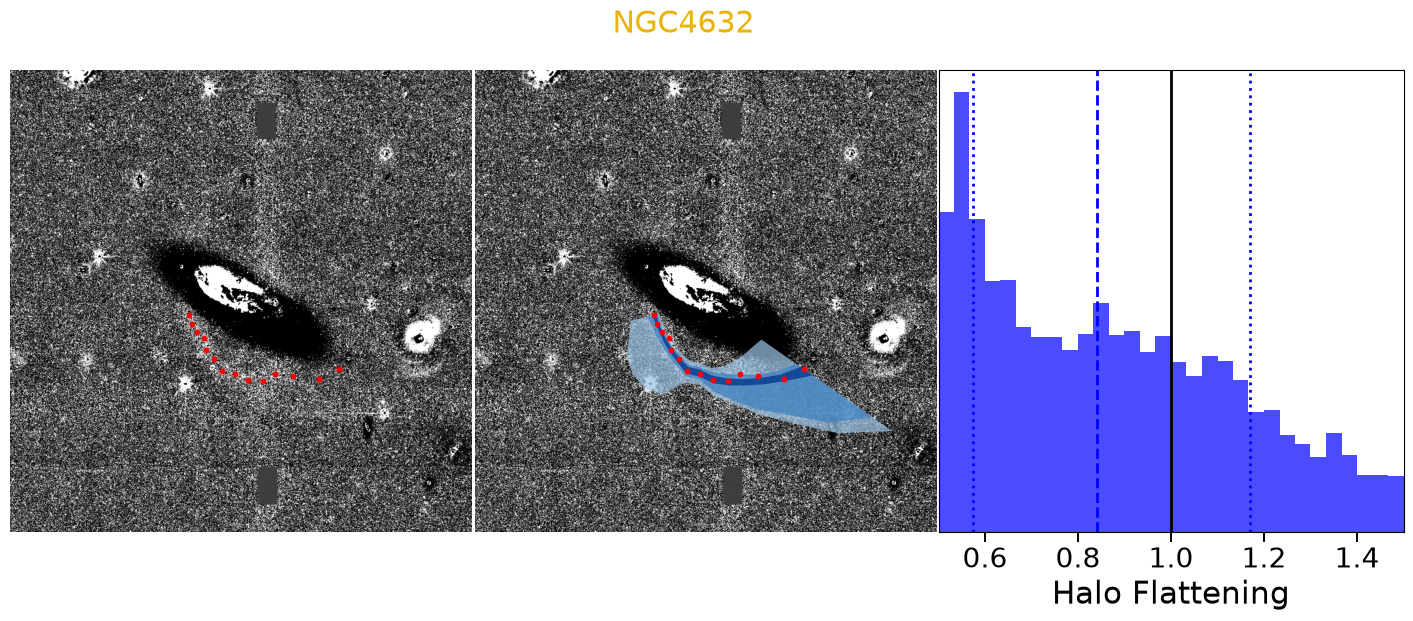}
    \end{minipage}
    \begin{minipage}{0.48\textwidth}
        \centering
        \includegraphics[width=\linewidth]{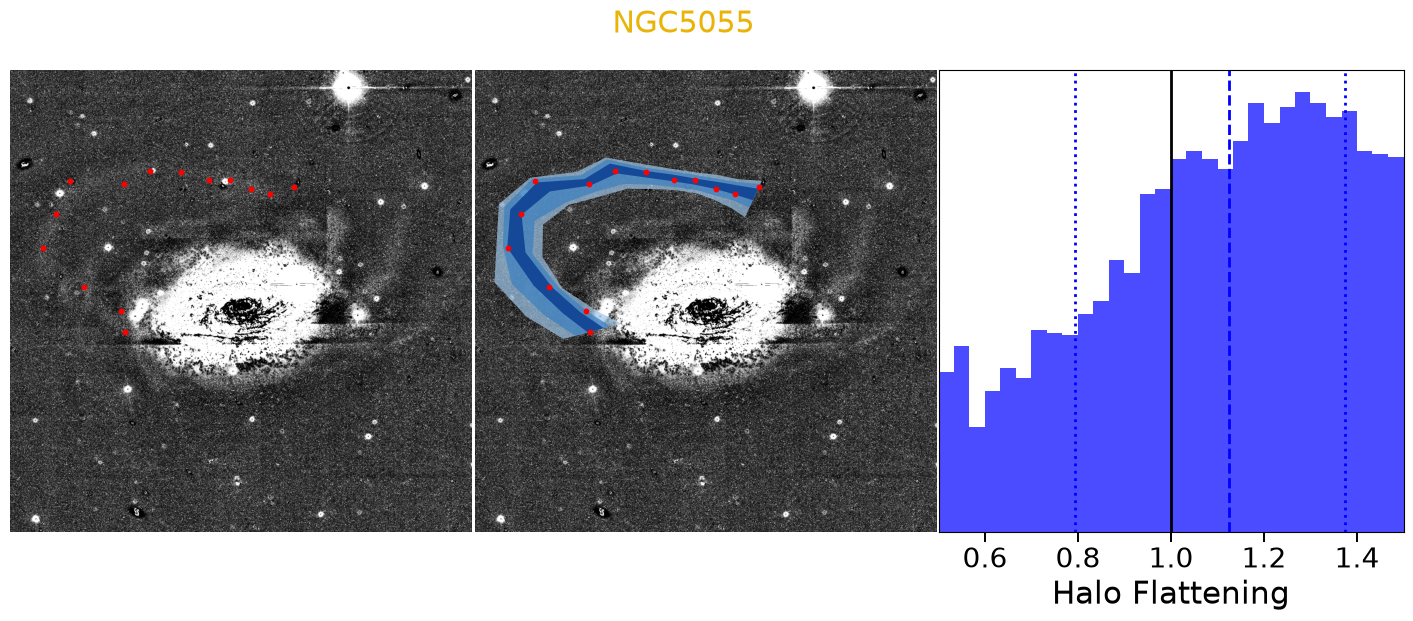}
    \end{minipage}

    \begin{minipage}{0.48\textwidth}
        \centering
        \includegraphics[width=\linewidth]{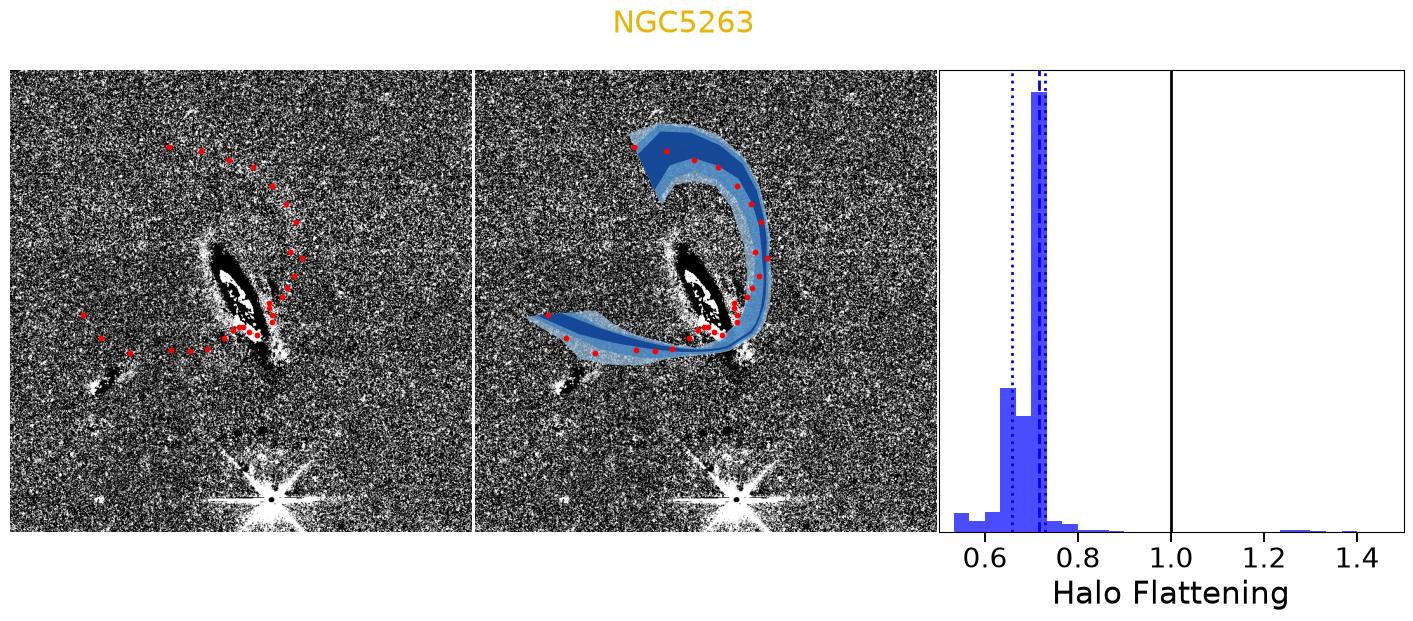}
    \end{minipage}
    \begin{minipage}{0.48\textwidth}
        \centering
        \includegraphics[width=\linewidth]{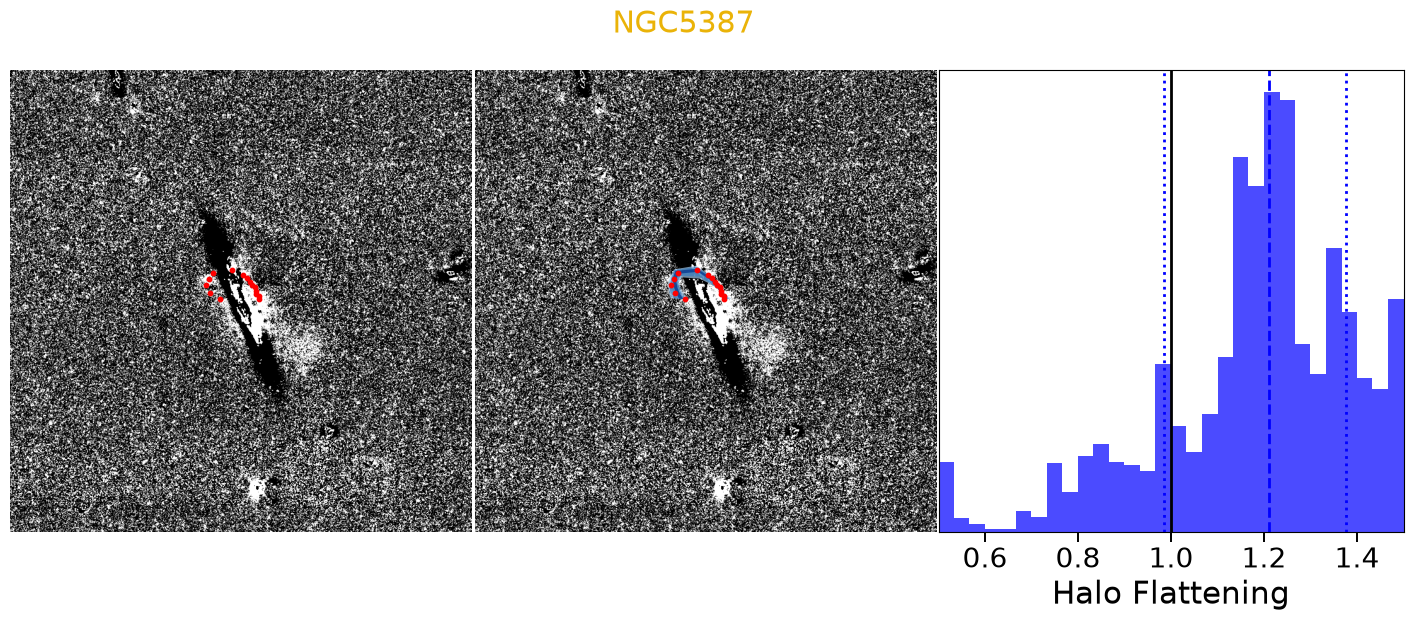}
    \end{minipage}
    \caption{Continuation of Figure~\ref{fig:example_individual} showing the remaining \textsc{STRRINGS} streams with the best fitting model shown in the middle panel and the posterior distribution of the halo flattening parameter shown in the right panel. The colour of each stream name indicates its subsample classification: goldenrod for the \emph{gold} sample and red for the \emph{bronze} sample.}
    \label{fig:AppB1}
\end{figure*}

\begin{figure*}
    \centering
    
    \begin{minipage}{0.48\textwidth}
        \centering
        \includegraphics[width=\linewidth]{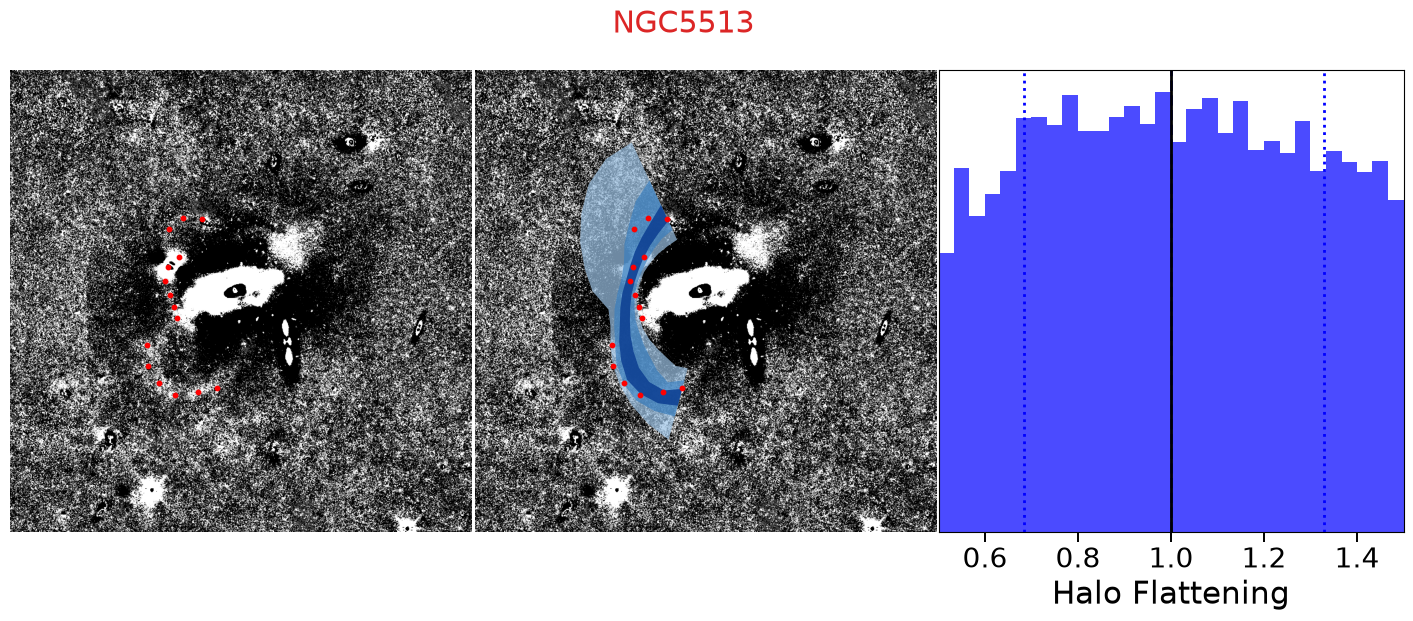}
    \end{minipage}
    \begin{minipage}{0.48\textwidth}
        \centering
        \includegraphics[width=\linewidth]{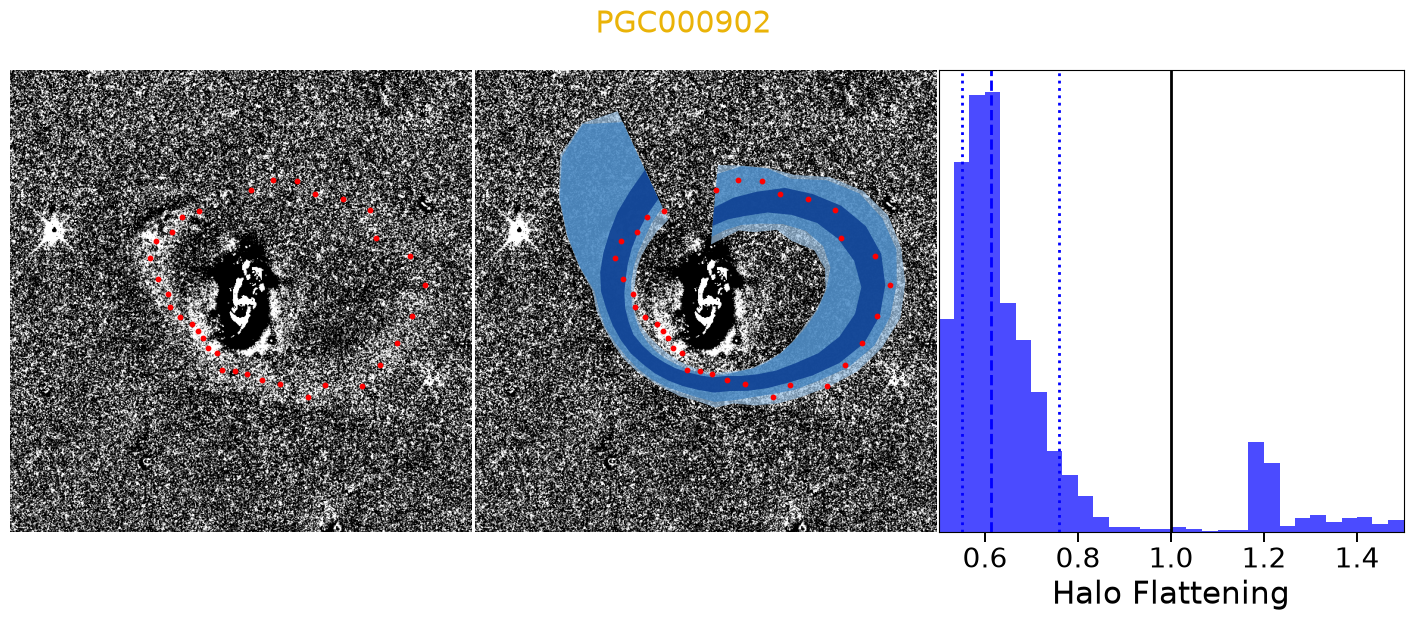}
    \end{minipage}

    \begin{minipage}{0.48\textwidth}
        \centering
        \includegraphics[width=\linewidth]{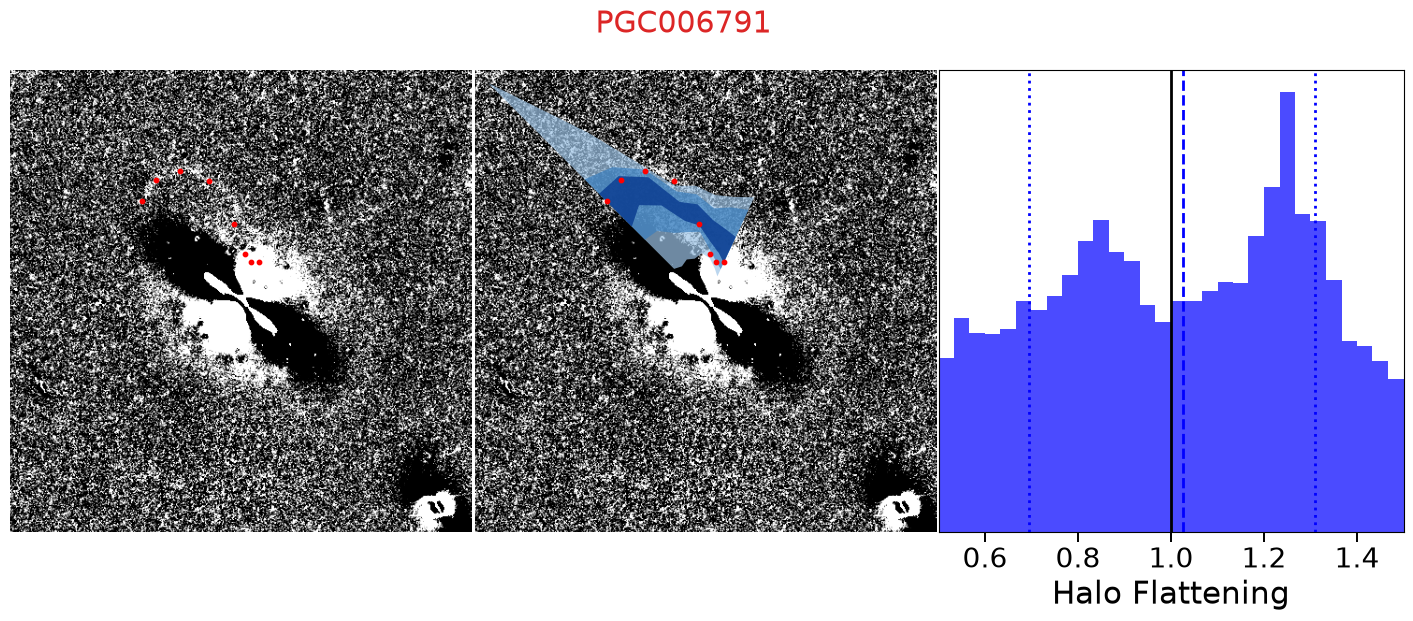}
    \end{minipage}
    \begin{minipage}{0.48\textwidth}
        \centering
        \includegraphics[width=\linewidth]{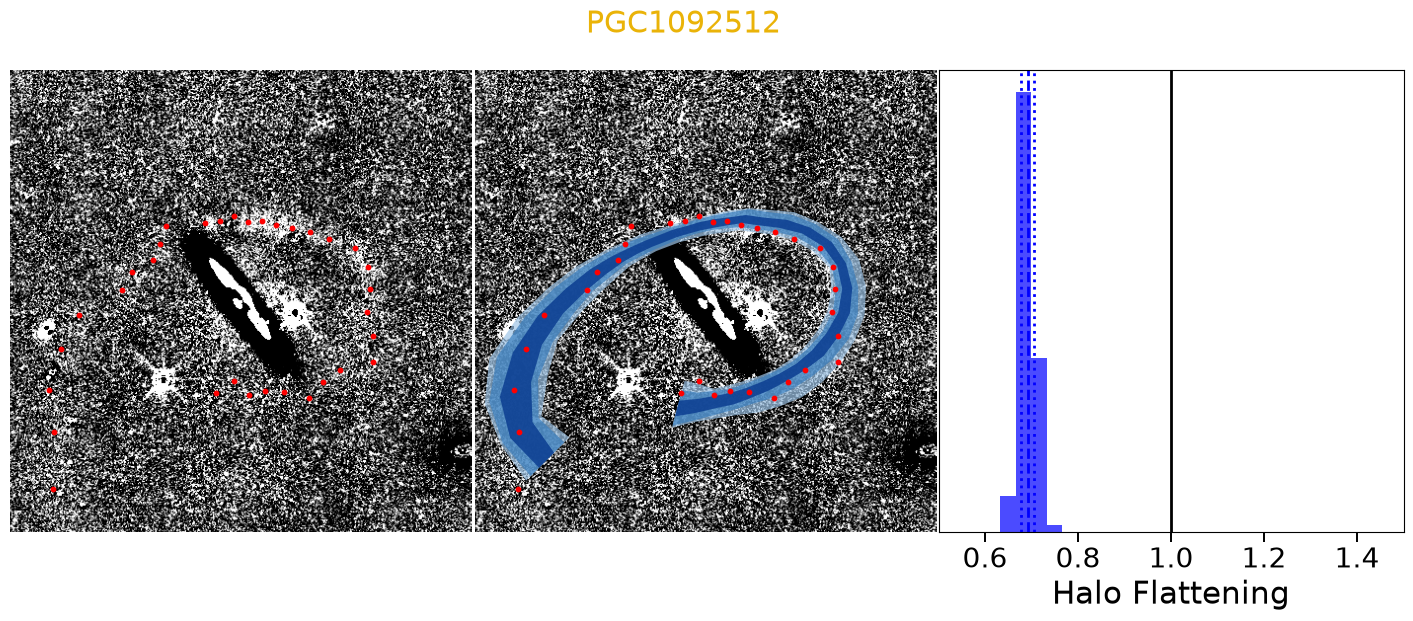}
    \end{minipage}

    \begin{minipage}{0.48\textwidth}
        \centering
        \includegraphics[width=\linewidth]{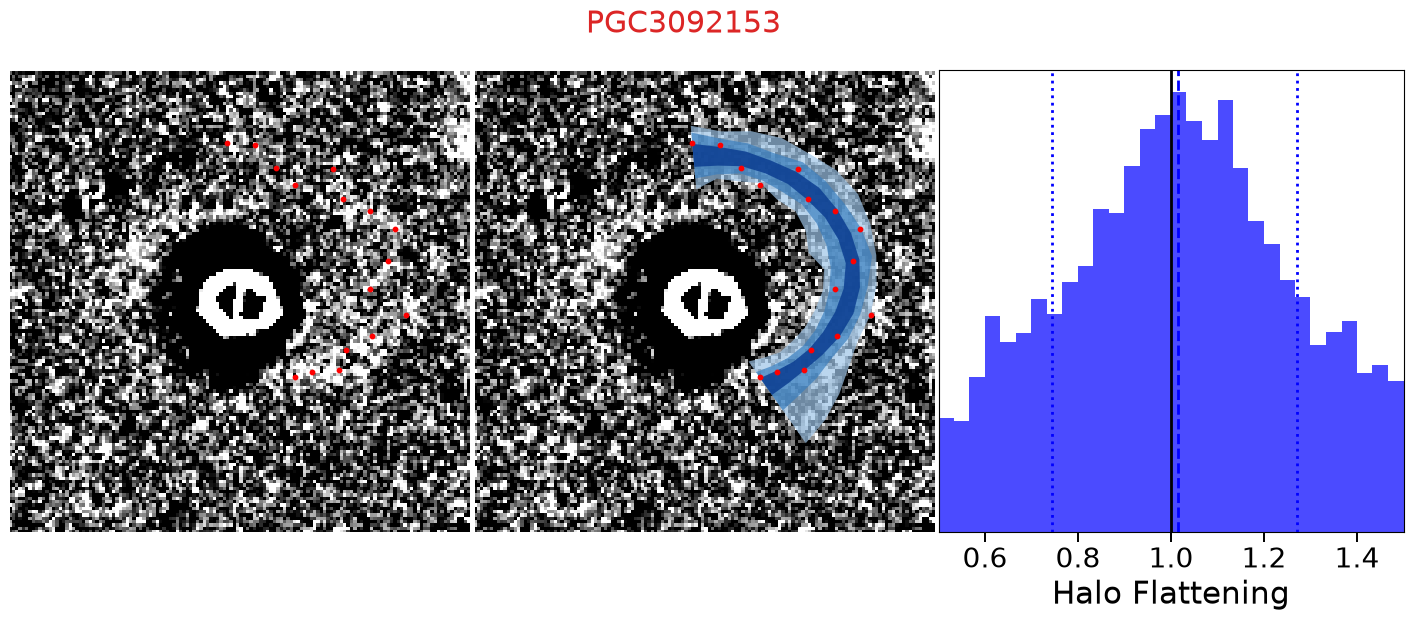}
    \end{minipage}
    \begin{minipage}{0.48\textwidth}
        \centering
        \includegraphics[width=\linewidth]{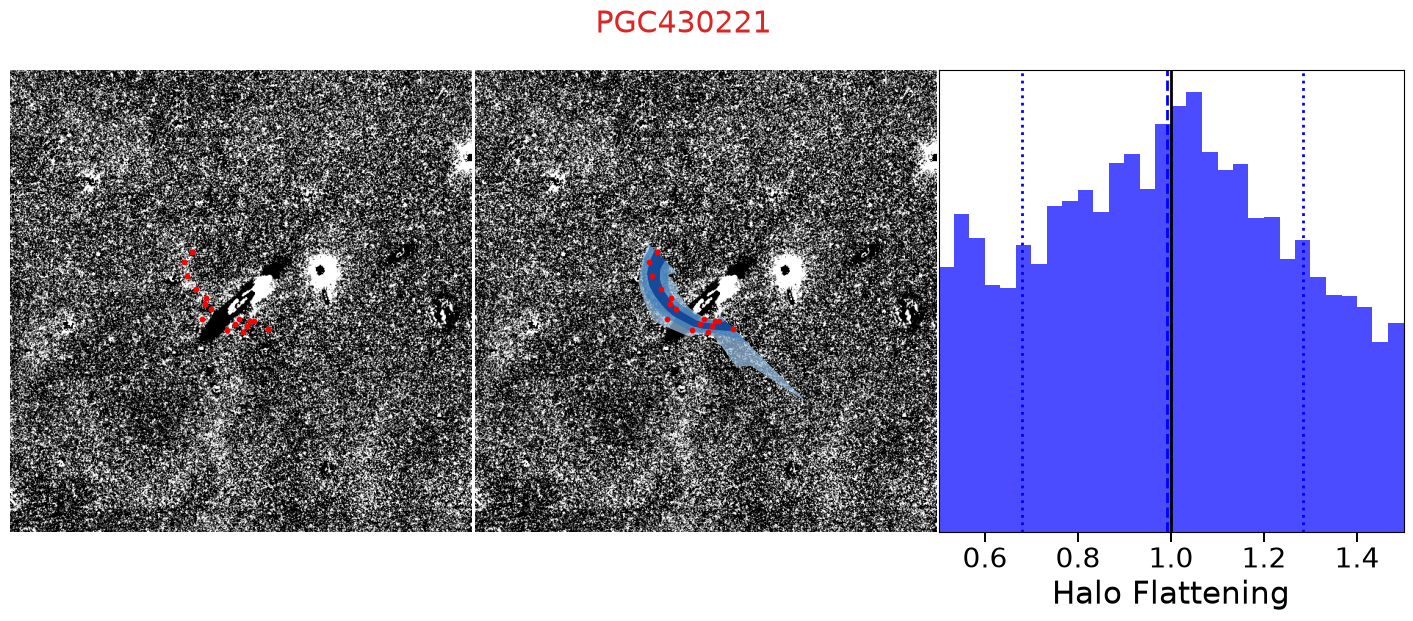}
    \end{minipage}

    \begin{minipage}{0.48\textwidth}
        \centering
        \includegraphics[width=\linewidth]{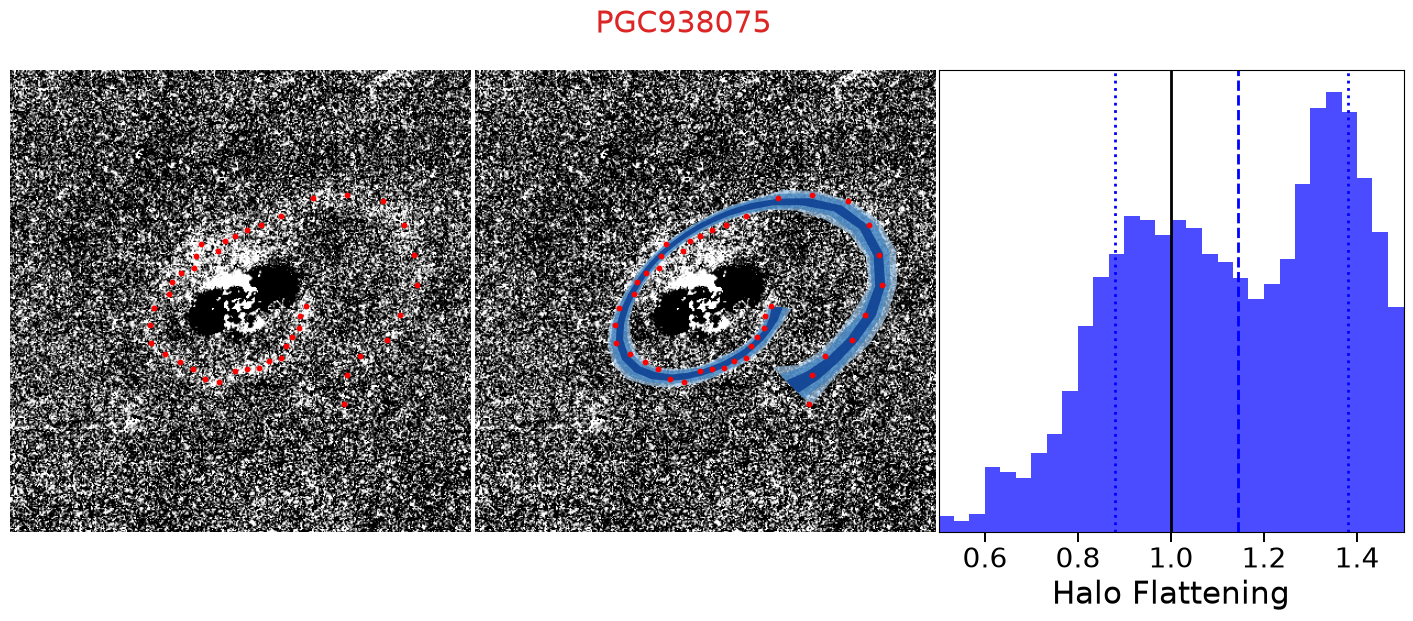}
    \end{minipage}
    \begin{minipage}{0.48\textwidth}
        \centering
        \includegraphics[width=\linewidth]{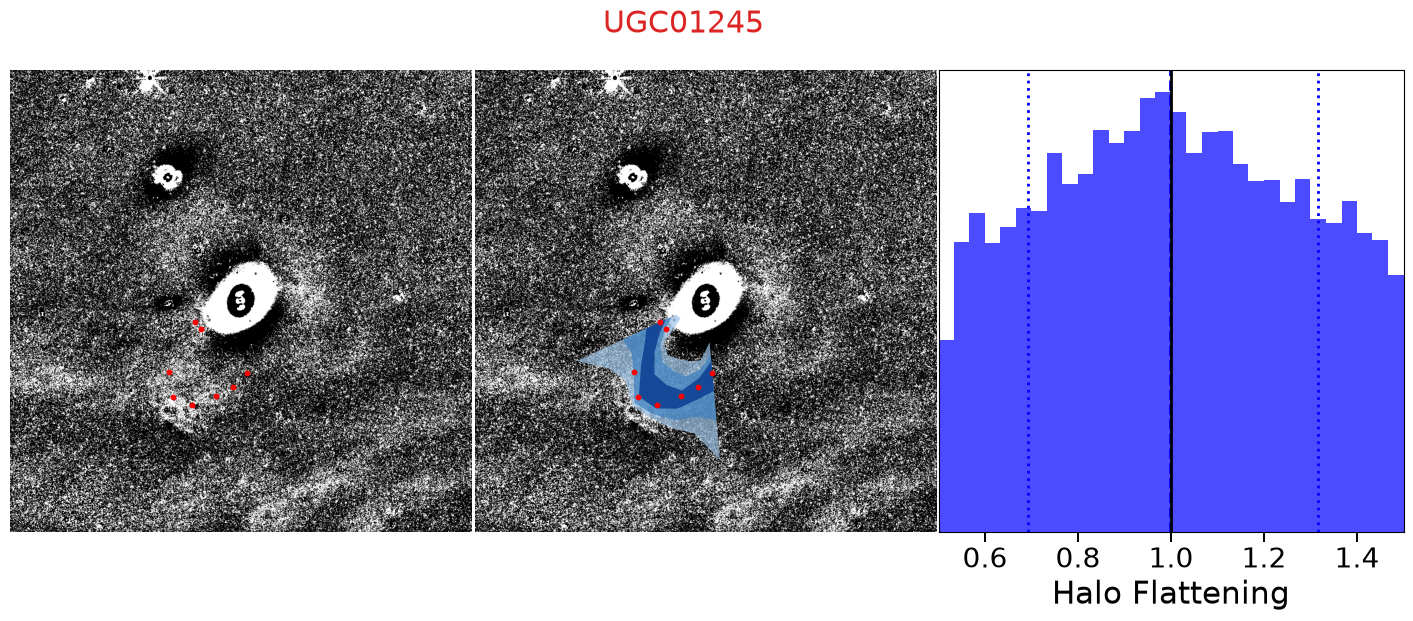}
    \end{minipage}

    \begin{minipage}{0.48\textwidth}
        \centering
        \includegraphics[width=\linewidth]{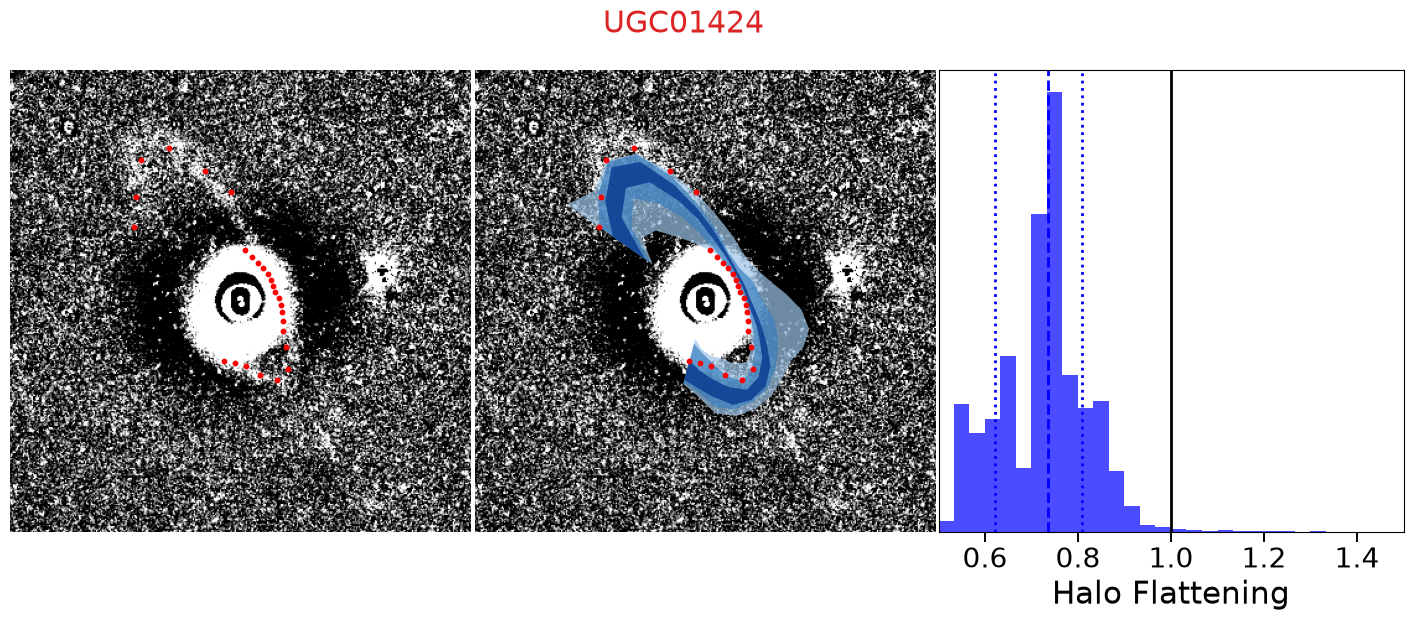}
    \end{minipage}
    \begin{minipage}{0.48\textwidth}
        \centering
        \includegraphics[width=\linewidth]{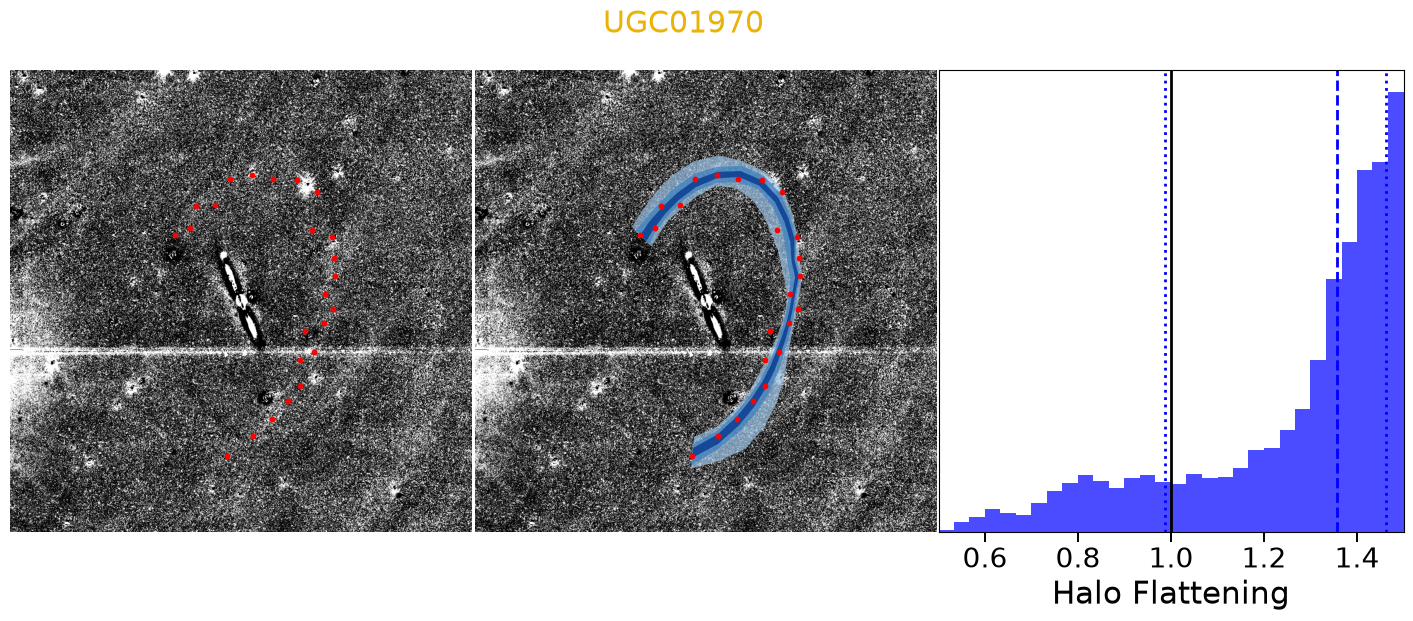}
    \end{minipage}

    \begin{minipage}{0.48\textwidth}
        \centering
        \includegraphics[width=\linewidth]{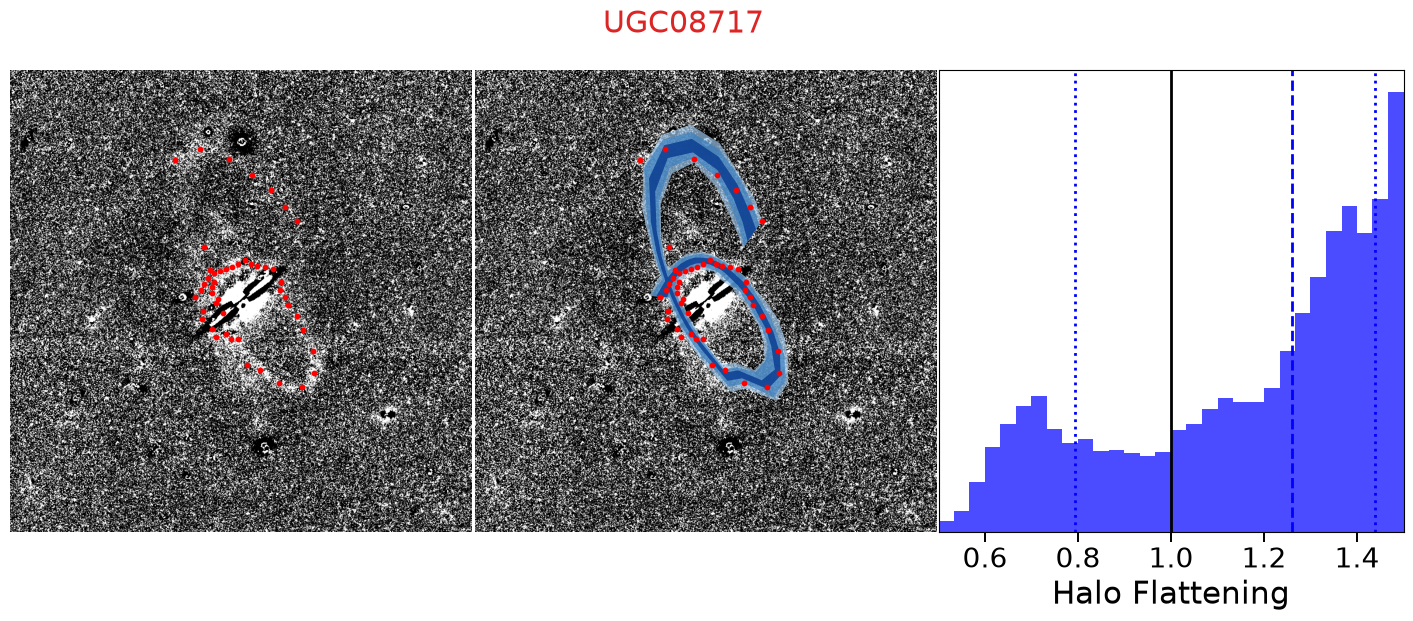}
    \end{minipage}
    \begin{minipage}{0.48\textwidth}
        \centering
        \includegraphics[width=\linewidth]{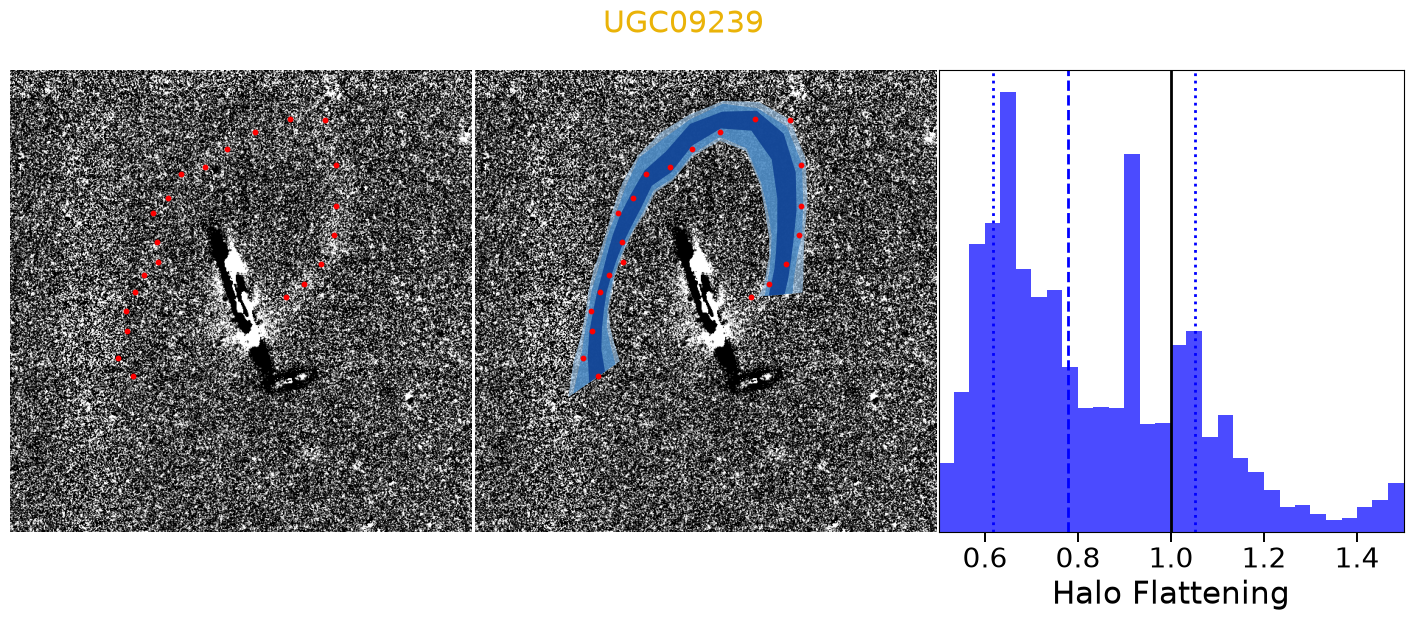}
    \end{minipage}
    \caption{Continuation of Figure~\ref{fig:AppB1}.}
    \label{fig:AppB2}
\end{figure*}

\section{Full posteriors of example fits}\label{app:corner}

\begin{figure*}
    \centering
    \includegraphics[width=0.92\linewidth]{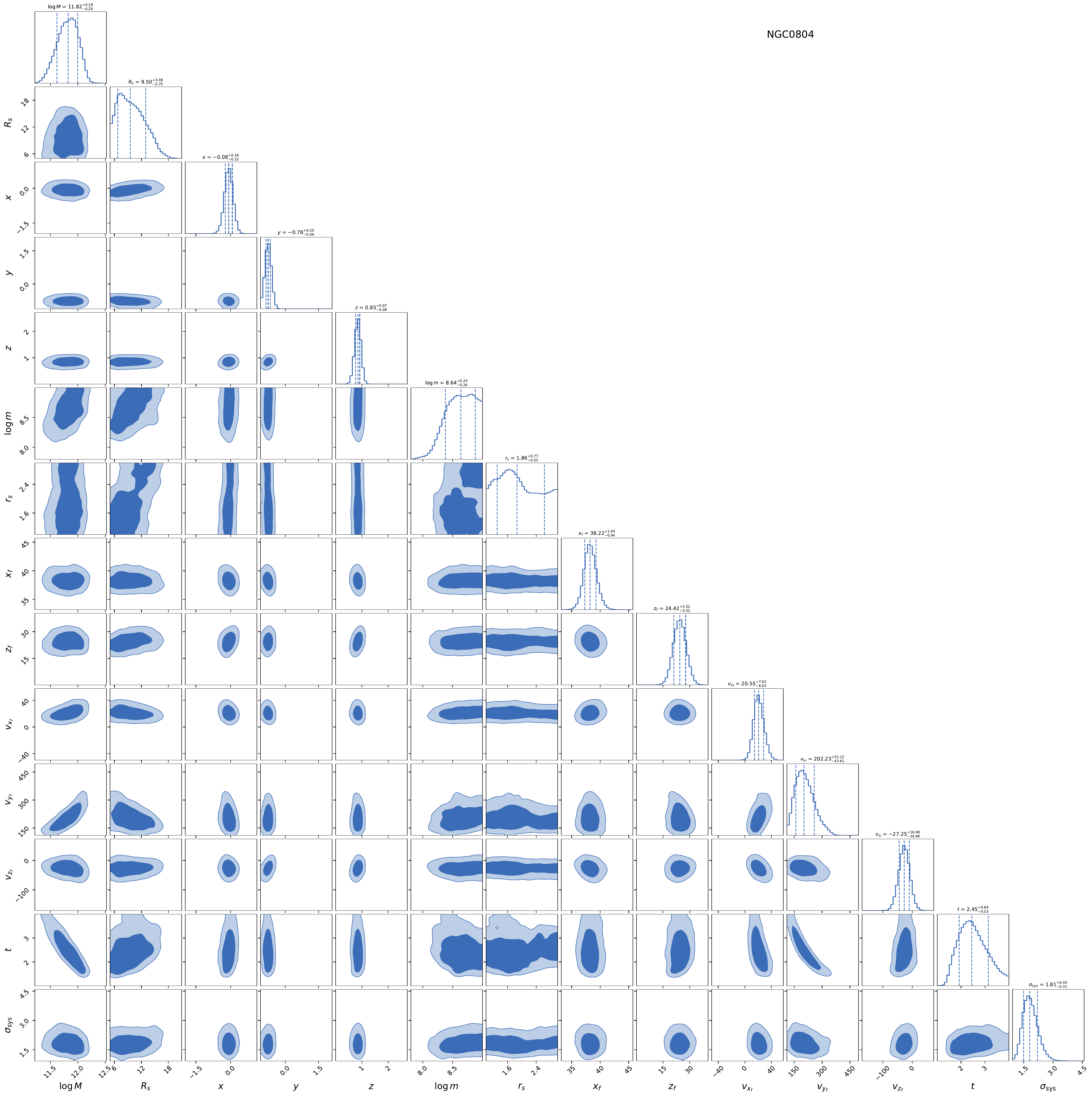}
    \caption{Full posterior distribution of all 14 model parameters for the example stream NGC0804, a member of the \emph{gold} sample: the host halo mass $\log M$ and scale radius $R_s$; the symmetry-axis orientation vector $(x,y,z)$, whose norm encodes the flattening $q$; the Plummer progenitor mass $\log m$ and scale radius $r_s$; the present-day phase-space coordinates $(x_f, z_f, v_{x_f}, v_{y_f}, v_{z_f})$; the integration time $t$; and the systematic scatter $\sigma_{\rm sys}$. NGC0804 is a very long stream, giving a high-information fit with tightly constrained orientation components (and hence $q$). Filled contours show the 68\% and 95\% credible regions (lightly smoothed for clarity); dashed lines and the panel titles give the 16th, 50th and 84th percentiles.}
    \label{fig:corner_example}
\end{figure*}

To illustrate a complete individual fit, Figure~\ref{fig:corner_example} shows the joint posterior distribution of all 14 model parameters for NGC0804, a member of the \emph{gold} sample (also shown in Figure~\ref{fig:example_individual}). It is one of our most informative fits: the stream spans a very long angular arc, which tightly constrains the symmetry-axis orientation vector $(x,y,z)$ and hence the halo flattening $q$. The diagonal panels show the marginal posteriors and the off-diagonal panels the pairwise correlations between parameters. Several physical degeneracies are apparent -- most notably an anti-correlation between the host mass $\log M$ and the integration time $t$ (a more massive halo is compensated by a shorter integration to trace out the same track), together with correlations between $\log M$ and the progenitor mass $\log m$ and present-day velocity, while the progenitor scale radius $r_s$ is only weakly constrained. We stress, however, that this fit is not fully representative of the sample. Because only the projected, on-sky stream track is observed, the orientation posterior is in some cases \emph{multimodal} -- distinct three-dimensional halo orientations can project to nearly the same track -- and for shorter streams the posteriors, particularly on the orientation and hence on $q$, are considerably broader and far less well constrained than the example shown here.

\bsp	
\label{lastpage}
\end{document}